\documentclass[final,aps,preprint,nofootinbib,groupedaddress,showpacs]{revtex4b3}
\usepackage{natbib}
\usepackage{amsmath}%
\usepackage{hyperref}%
\usepackage{graphicx}
\newcommand{\noopsort}[1]{} 
 
\begin{document}

% Again put in by author for 1" binding
%\oddsidemargin 0.5in \evensidemargin 0in \textwidth 6in
% \topmargin 0.5in

\brokenpenalty=10000

\preprint{ASTRO-PH/0005003}

\title{An Exposition on Inflationary Cosmology}
\author{Gary Scott Watson}
\email[email:]{thewatsons@prodigy.net}
\thanks{Research partially supported by a grant from the University of North Carolina at Wilmington Committee for Undergraduate Research}
\affiliation{University of North Carolina at Wilmington\\
Department of Physics\\ 601 South College Road\\ Wilmington, NC
28403}
\date{\today}

\begin{abstract}
This paper is intended to offer a pedagogical treatment of
inflationary cosmology, which is accessible to undergraduates. In
recent years, inflation has become accepted as a standard scenario
making predictions that are testable by observations of the cosmic
background. It is therefore manifest that anyone wishing to pursue
the study of cosmology and large-scale structure should have this
scenario at their disposal. The author hopes this paper will serve
to `bridge the gap' between technical and popular accounts of the
subject.
\end{abstract}

\maketitle

%__________________ACKNOWLEDGEMENTS____________________________________
\begin{center}
\bf{Acknowledgments}
\end{center}

I would like to thank:

\begin{itemize}
\item Gabriel Lugo (UNCW) and Brian Davis (UNCW) for being on my examining
committee, reviewing my paper, and years of supporting my
education,

\item Paul Frampton and the UNC Cosmology and Particle Theory Group
for inviting me for weekly visits and for many useful discussions,

\item Wayne Hu of the Institute for Advance Study at Princeton and
Sean Carroll of the University of Chicago for their contributed
images and pedagogical assistance,

\item Andrei Linde of Stanford University for proofing the paper and
offering several useful suggestions and references.
\end{itemize}

I would especially like to thank Edward Olszewski (UNCW) and
Russell Herman (UNCW) for their tireless efforts, a list of which
would occupy more than the space allotted.
% ______________________________END ACKNOWLEDGEMENTS_________________________

\pagebreak
\vspace*{2in}
\begin{center} {\Large{{\em To my wife, Kara}}}
\end{center}
\newpage
\tableofcontents
\newpage
%------------------------------------Begin Introduction-----------------------
\section{Introduction}

The Standard Model of Cosmology has successfully predicted the
nucleosynthesis of the light elements, the temperature and
blackbody spectrum of the cosmic background radiation, and the
observed redshift of light from galaxies which suggests an
expanding universe. However, this model can not account for a
number of initial value problems, such as the flatness and
monopole problems. Inflationary cosmology resolves these concerns,
while preserving the successes of the Big-Bang model. Inflation
was originally introduced for this reason and its motivation
relied on predictions from particle theory.  In more recent times,
inflation has been abstracted to a much more general theory.  It
continues to resolve the initial value problems, but also offers
an explanation of the observed large-scale structure of the
universe.

In this paper, the fundamentals of modern cosmology for an
isotropic and homogeneous space-time, which is naturally motivated
by observation, will be reviewed.  The Friedmann equations are
derived and the consequences for the dynamics of the universe are
discussed. A brief introduction to the thermal properties of the
universe is presented as motivation for a discussion of the
horizon problem. Moreover, other issues suggesting a more general
theory are presented and inflation is introduced as a resolution
to this conundrum.

Inflation is shown to actually exist as a scenario, rather than a
specific model.  In the most general case one speaks of the
inflaton field and its corresponding energy density. Models of
inflation differ in their predictions and the corresponding
evolution of an associated inflaton field can be explored in a
cosmological context.  The equations of motion are cast in a form
that makes observational consequences manifest. The slow-roll
approximation (SRA) is discussed as a more tractable and plausible
evolution for the inflaton field and the slow-roll parameters are
defined. Using the SRA, inflation predicts a near-Gaussian
adiabatic perturbation spectrum resulting from quantum
fluctuations in the inflaton field and the DeSitter space-time
metric.  These result in a predicted power spectrum of gravity
waves and temperature anisotropies in the cosmic background, both
of which will be detectable in future experiments.

Inflation is shown to be a rigorous theory that makes concise
predictions in regards to a needed inflaton potential at the
immediate Post-Planck or perhaps even the Planck epoch ($\sim
10^{-43}$s). This offers the exciting possibility that inflation
can be used to predict new particle physics or serve as a
constraint for phenomenology from theories such as Superstring
theory.
%------------------------------------End Introduction-------------------------
\newpage
%------------------------------------Begin Section One------------------------
\section{Standard Cosmology}

\subsection{The Cosmological Principle}

The Cosmological Principle (CP) is the rudimentary foundation of
most standard cosmological models.  The CP can be summarized by
two principles of spatial invariance.  The first invariance is
isomorphism under translation and is referred to as homogeneity.
An example of homogeneity can be seen in a carton of homogeneous
milk. The milk or liquid, looks the same no matter where one is
located within it.  In the realm of cosmology, this corresponds to
galaxies being uniformly distributed throughout the universe. This
uniformity would be independent of the location one chooses to
make the observations.  Thus, a translation from one galaxy to
another would leave the galactic distribution invariant
(invariance under translation).

The next element of the CP is perhaps more difficult to be
realized physically.  This invariance is isomorphism under
rotation and is referred to as isotropy.  A simple way of
visualizing isotropy is to say that direction, such as North or
South, can not be distinguished.  For example, if one were
constrained to live on the surface of a uniform sphere, there
would be no geometrical method to distinguish a direction in
space. Although, as soon as features are introduced on the sphere
(such as land masses or cracks in the surface of the sphere), the
symmetry is lost and direction can be established. This fact gives
a clue that isotropy, as you might have guessed, is closely
related to homogeneity.

The concepts of homogeneity and isotropy may appear contradictory
to local observation.  The Earth and the solar system are not
homogeneous nor isotropic.  Matter clumps together to form objects
like galaxies, stars, and planets with voids of near-vacuum in
between. However, when one views the universe on a large scale,
galaxies appear `smeared out' and the CP holds.

Experimental proof of isotropy and homogeneity has been approached
using a number of methods.  One of the most convincing
observations is that of the Cosmic Background Radiation (CBR). In
the standard Big Bang model, the universe began at a singularity
of infinite density and infinite temperature.  As the universe
expanded it began to cool allowing nucleons to combine and then
atoms to form.  About 300,000 years after the Big Bang, radiation
decoupled from matter, allowing it to `escape' at the speed of
light.  This radiation continues to cool to the present day and is
observed as the CBR.  As we will see, observations of the CBR
gives a picture of the mass distribution at around 300,000 years.
The temperature of the CBR, first predicted theoretically in the
1960's by Alpher and Herman at $5K$ \citep{alpher}, and Gamow at a
higher $50K$ \citep{Gamow}, was not taken seriously. A later
prediction by Dicke, et. al.  \citep{dicke} yielded $\approx 3 K$,
but as Dicke and colleagues set out to measure this remnant
radiation, they found someone had already made this measurement.
Dicke remarked, ``Well boys, we've been scooped''
\citep{partridge}.

\begin{figure}[]
\includegraphics[totalheight=5.0 in,keepaspectratio]{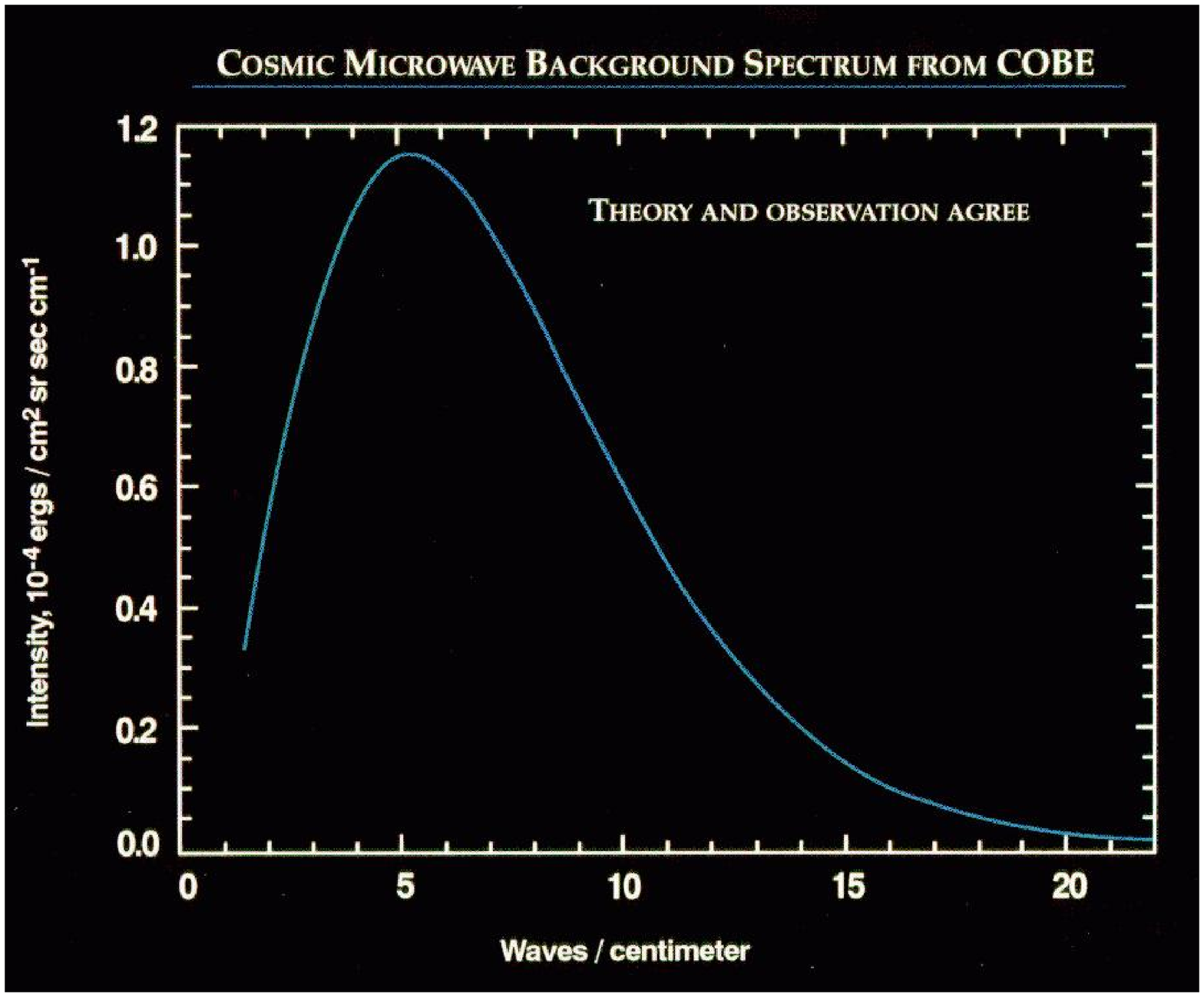}
\caption{The cosmic background spectrum is that of a near perfect
blackbody as predicted by theory.  The above graph represents the
most recent observations by a number of collaborations.  Graph
courtesy of COBE Project \citep{cobe}} \label{cmbspectrum}
\end{figure}

The first successful measurement of the CBR was made in 1964 by
Penzias and Wilson, two scientists working on a satellite
development project for Bell Labs \citep{partridge}. Their
measurements revealed that the CBR was characteristic of a
black-body with a corresponding temperature of around $3K$ as
illustrated in Figure (\ref{cmbspectrum}) \citep{partridge}. The
measured wavelengths were on the order of $7.35$ cm, corresponding
to the microwave range of the electromagnetic spectrum. The CBR in
this range is referred to as the Cosmic Microwave Background
(CMB)\footnote{The significance in making this distinction will
manifest itself later, but it is worth noting that other
backgrounds are measurable and offer further evidence of the CP.}.

Another important observation of Penzias and Wilson is the fact
that the CMB is uniform (homogeneous) in all directions
(isotropic). Thus, the CMB offers an experimental proof of the
isotropy and homogeneity of the universe.

Because of its importance, further measurements of the CBR have
been carried out. One such project named COBE, for Cosmic
Background Explorer, in $1989$, measured the CBR to have a
temperature of $2.73$ K and a distribution that is isotropic to
one part in $10^{5}$ \citep{turner}. COBE also has the distinction
of being the first satellite dedicated solely to cosmology. Future
measurements will be made by dedicated satellites like COBE, but
these satellites will have much higher angular resolution. They
are planned to be launched around the beginning of the century.

Balloon born experiments have been able to measure the background
spectrum with greater resolution than COBE and the preliminary
results seem to favor the type of spectrum predicted by the
inflationary scenario, to be discussed later
\citep{Melchiorri:1999br},\citep{Lange:2000iq}. Several satellite
projects are planned, MAP, for Microwave Anisotropy
Probe\footnote{For more info see: http://map.gsfc.nasa.gov/} will
be launched at the end of this year by NASA and another named the
PLANCK Explorer is planned for launched by the European Space
Agency\footnote{http://astro.estec.esa.nl/SA-general/Projects/Planck/}
around the year 2006. The accurate measurement of the CBR offers
an observational test of cosmological models, as well as, the CP.

In addition to these benefits of CBR observations, the CBR can
also be used to setup a Cosmic Rest Frame (CRF).  This concept is
reminiscent to the ideas of Ernst Mach.  One chooses a reference
frame to coincide with the Hubble expansion, i.e., with the motion
of the average distribution of matter in the universe. It is
convenient to define our coordinates in this frame to save
confusion in measurements such as the expansion of spacetime and
the Hubble Constant; however, these coordinates are in no way
`absolute' coordinates. Using the CBR to define the CRF and taking
galaxies as the test particles of the model serves to greatly
simplify the dynamics in an expanding universe. The CRF is used to
ease calculations and make the interpretation of the dynamics of
an expanding universe more tractable.

The current and proposed measurements of the CBR offer a
convincing test of the homogeneity of space.  Measurements of the
temperature of the CBR are uniform to one part in $10^{5}$.  This
suggests the universe is homogeneous and isotropic to a high
degree of accuracy. However, since this measurement is taken from
our (the Earth's) vantage point, one can not assume the same
conclusion from another vantage point.  This can be remedied by
considering how the CBR is related to the distribution of matter
at the time the photons of the CBR decoupled.  This offers a `snap
shot' of the inhomogeneities in the density of the universe.  If
these regions contained more inhomogeneity, galaxies would not be
visible today. This idea will be discussed in more detail later;
as an alternative one can introduce the Copernican Principle (CP).

The CP states that no observers occupy a special place in the
universe. This appears to be a favorable prediction, based on the
evidence above, as well as lessons coming from the past. For
example, the correct model of the solar system was not realized
until humans realized they were not the center of the solar
system.  This may be a bit humbling to the human ego, but the
Copernican Principle, along with homogeneity and isotropy, serve
to greatly simplify the number of possible cosmological models for
the universe.  Later, it will be seen that homogeneity follows
naturally from inflation. If the universe went through a brief
period of rapid expansion, the fact that galaxies exist at all
will be a necessary and sufficient condition for a homogeneous
universe.

There is also the proposal for cosmic `no-hair' theorems.  These
theorems are similar to the `no-hair' proposal of black holes,
which predict that any object that contains an event horizon will
yield a Schwartzschild spherically symmetric solution at the
singularity. The Big-Bang singularity is no exception, and the
event horizon is the Hubble distance to be explored in sections to
come. For now, experiment suggests that it is safe to assume the
Copernican Principle is valid.

Below is a brief descriptive summary of observational methods for
testing the CP:

\begin{itemize}

\item {\em Particle Backgrounds} -- These observations represent the
strongest argument for isotropy and homogeneity.  As the universe
evolved it cooled allowing various particle species to become
`frozen out', meaning that the particles were freed from
interactions. Photons, for example, became frozen out at the time
of decoupling and are visible today as the CBR.  These backgrounds
serve as an important experimental test for predictions by various
cosmological models.

\item {\em The Observed Hubble Law} -- This law states that the farther away a galaxy
is, the faster it will be observed to recede\footnote{One must be
careful here, as we will see the spacetime between the galaxy and
us is actually what is expanding, the galaxy itself is not really
receding.}. This phenomena is observed through a redshift of the
light coming from the galaxy and will be described in a later
section. The observed redshift, first witnessed by Edwin Hubble
was the first indication that the universe obeys the CP.

\item {\em Source Number Counts} -- Of all methods this is the most
uncertain at this time.  This method requires collecting light
from galaxies and inferring whether `clustering' occurs.  One
debate over the accuracy of such methods is based on the idea that
most matter in the universe might be of a non-luminous type, the
so-called Dark Matter. Another problem is that current technology
does not allow observations at distances far enough to get a good
sample of the population. However, this technique shows promise
for the future, and the SLOAN\footnote{http://www.sdss.org/}
Digital Sky Survey is a current project that will map in detail
one-quarter of the entire sky, determining the positions and
absolute brightness of more than $100$ million celestial objects.
It will also measure the distances to more than a million galaxies
and quasars.

\item {\em Inflation} -- Although it is premature at this point to
discuss observational consequences of inflation, it will be shown
that inflation predicts small perturbations in the universe that
result in the large-scale structure observed today.  It will be
shown that if these perturbations were too large then the
structure we observe today would not be possible.  Thus, if
inflation can be proved through observation, it would imply the
universe must have been very homogeneous at the time of
decoupling.
\end{itemize}

The established concepts of the CP aid in simplification of
cosmological models, but a further simplification can be made by
invoking the Perfect Cosmological Principle.  This principle
differs from the previous one in that it assumes temporal
homogeneity and isotropy.  This would imply a static universe, for
if the universe were expanding or contracting it would not look
the same now, as it did in the past. However, one exception that
will prove important later is the case of a (anti or quasi)
DeSitter Space.  By the observations of Edwin Hubble and the
theoretical work by Lama\^\i tre\footnote{Lama\^\i tre will not be
mentioned further but it is worth noting that his work and
persistence, backed by the experimental efforts of Hubble, were
instrumental in convincing Einstein that the universe was indeed
expanding.  After this persuasion, Einstein was quoted as saying
this was the biggest mistake of his career \citep{hawley}.} it was
shown that the expansion of the universe is an accurate
assumption.  CP models further suggest that a static universe
would be as stable as a pencil standing on its end. Thus, the
Perfect Cosmological Principle does not appear to be an acceptable
assumption within the standard model \citep{hawking}\footnote{This
is not totally correct. In some space-times, such as anti-DeSitter
space, there exists temporal homogeneity.  For a rigorous
treatment of such space-times consult \citep{hawking}.}.

The last element to be discussed concerning the CP is the Weyl
Postulate.  This postulate formally states that, ``the world lines
of galaxies designated as `test particles' form a 3-bundle of
nonintersecting geodesics orthogonal to a series of spacelike
hypersurfaces'' \citep{narlikar}.  In other words, the geodesics
on which galaxies travel do not intersect.  This adds another
symmetry to the picture of the expanding universe allowing
simplification of the spacetime metric and the Einstein equations.

\subsection{The Expanding Universe}

In the mid-twenties, Edwin Hubble was observing a group of objects
known as spiral nebulae\footnote{It would later be found that most
of these nebula were in fact galaxies \citep{hawley}.}. These
nebulae contain a very important class of stars known as Cepheid
Variables. Because the Cepheids have a characteristic variation in
brightness \citep{bergstrom}, Hubble could recognize these stars
at great distances and then compare their observed luminosity to
their known luminosity. This allowed him to compute the distance
to the stars, since luminosity is inversely proportional to the
square of the distance \citep{bergstrom}. The intrinsic, or
absolute, luminosity is calculated from simple models that have
been commensurate with observations of near Cepheids.

When Hubble compared the distance of the Cepheids to their
velocities (computed by the redshift of their spectrum) he found a
simple linear relationship,
\begin{equation}
\label{hubble law} {\vec V_{H}}=H \vec{r},
\end{equation} where ${\vec V_{H}}$ is the velocity of the galaxy, $H$ is the
so-called Hubble Constant, and $\vec{r}$ is the displacement of
the galaxy from the Earth. It will be shown later that the Hubble
constant is not actually a constant, but can be a function of time
depending on the chosen model.  The standard notation is to adopt
$H_{0}$ as the `current' observed Hubble parameter, whereas
$H=H(t)$ is referred to as the Hubble constant. The current
accepted value of the Hubble parameter is,
\begin{equation} \label{hubble const} H_{0}=100 h_{0} \; \mbox{km}
\cdot \mbox{s}^{-1} \cdot \mbox{Mpc}^{-1}\;\;\;\;\;
\mbox{where}\;\;0.5 < h_{0} < 0.8.
\end{equation} The unit of length, Mpc, stands for Megaparsec\footnote{$1 \; Mpc =
10^{6} \; parsecs \approx 3 \; light years \approx 3 \times
10^{16} \; meters.$ \\A parsec is the distance to an object that
has an angular parallax of $1^{\circ}$ and a baseline of 1 A.U.
For more on Observational Astronomy see \citep{filippenko}. }.

Hubble's interpretation of his data was crucial in helping
determine the correct model for the universe.  Hubble had found
that the galaxies, on average, were receding away from us at a
velocity proportional to their distance from us (\ref{hubble
law}).  This suggests a homogeneous, isotropic, and expanding
universe.  By this finding, the choices of cosmological models
became greatly restricted.

Perhaps it is worth mentioning that the above analysis by Hubble
is not quite as easily done as one might think.  One factor that
must be considered in the calculation of the Hubble velocity field
(\ref{hubble law}) is the concept of peculiar velocity.  This is
the name given to the motion of a galaxy, relative to the CRF, due
to its rotation and motion as influenced by the gravitational pull
of nearby clusters. This speed, $v_{p} \leq \pm 500 \; \mbox{km}
\cdot \mbox{s}^{-1}$, can be neglected at far distances where the
Hubble speed, ${V_{H}} \gg  500 \; \mbox{km} \cdot \mbox{s}^{-1}$.
Thus, when Hubble conducted his survey most of the nebulae were
too near to rule out an effect by the peculiar velocity. As a
result, Hubble found $H_{0} \approx 500 \; \mbox{km} \cdot
\mbox{s}^{-1} \cdot \mbox{Mpc}$, much greater than the value
obtained today from surveys of type Ia
supernovae\footnote{Supernova Ia, like Cepheid Variables, have a
known `signature' and can therefore be used as `Standard Candles',
but unlike the Cepheids, supernovae are much more luminous and can
therefore be seen at much greater distances \citep{richtler}.}.

\subsubsection{The Hubble Law and Particle Kinematics} 
\begin{figure}[]
\includegraphics[totalheight=4.0 in,keepaspectratio]{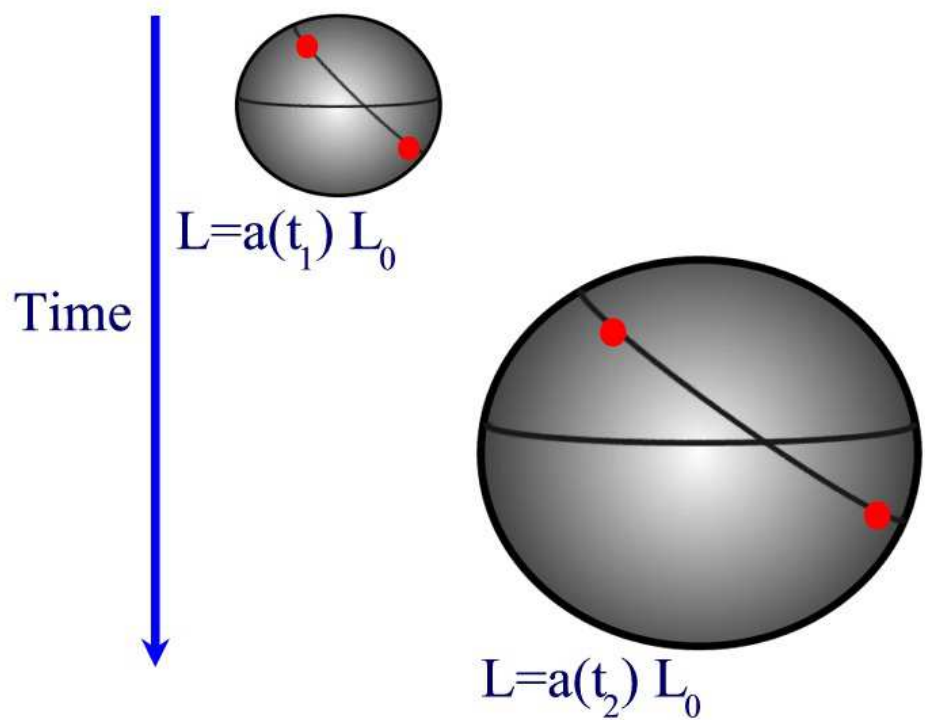}
\caption{As the universe expands the distance between galaxies
(test particles) increases proportional to the scale factor,
$L(t)=a(t) L_{0}$.} \label{hub1}
\end{figure}

The Hubble
law (\ref{hubble law}) is a direct result of the CP. Consider the
expansion of the universe, which must occur in a homogeneous and
isotropic manner according to the CP.  The expansion can be
visualized with the analogy of a balloon with a grid painted on
it.  Of course this should not be taken literally, since the
spatial extent of the universe is three dimensional. Think of the
grid as a network of meter sticks and clocks at rest with respect
to the Hubble expansion, which corresponds to the Cosmic Rest
Frame (CRF) mentioned earlier. Due to the expansion, two
particles\footnote{Remember that when one speaks of a cosmological
model, the test particles are galaxies.} initially separated by a
distance $l_{0}$, will be separated by a distance $l(t)=a(t)l_{0}$
at some later time $t$, see Figure (\ref{hub1}). Because of the
CP, the function $a(t)$, known as the scale factor, can only be a
function of time. From this relation, the speed of the observers
relative to each other is,

$$v(t)=\frac{dl}{dt}=\dot{a} l_{0}= \Big( \frac{\dot{a}}{a}
\Big)l(t) = H(t) l(t),$$ where $\dot{a}$ is the time derivative of
the scale factor.

From this derivation of the Hubble law, it becomes manifest that
the Hubble Constant can depend on time.  In this new way of
defining $H(t)=\dot{a}(t)/a(t)$, $H(t)$ measures the rate of
change of the scale factor, $a(t)$, and offers a way to link
observations (like Hubble's) with a proposed model using the scale
factor.  For Hubble's observations, the distance $l(t)$ was small
and $H(t)$ could be estimated by a linear relation yielding
equation (\ref{hubble law}).

To understand how particles `come to rest' in the CRF, consider a
particle starting out with a peculiar velocity $v_{p}  \ll c$. The
particle passes a CRF observer ($O_{1}$) at time $t$ and travels a
distance $dl=v_{p} \: dt$.  At this time the particle passes
another CRF observer ($O_{2}$), who has a velocity $dv=H dl=H
v_{p}\: dt$ relative to $O_{1}$.  $O_{2}$ measures the particle's
peculiar velocity as, $v_{p}(t+dt) = v_{p}(t)- dv $. This shows
that the peculiar velocity satisfies the equation of motion,
\begin{equation}
\label{peculiar velocity} \frac{dv_{p}}{dt}=-\frac{dv}{dt}= -H
v_{p}=- \Big( \frac{\dot{a}}{a} \Big) v_{p}.
\end{equation}
Solving this differential equation yields, $$v_{p} \propto
\frac{1}{a} \;.$$ This indicates that the peculiar velocity
decreases as the scale factor increases.  Indicating that as the
universe expands, particles with peculiar velocities tend to go to
zero meaning they `settle' into the CRF.

\subsubsection{The Robertson Walker Metric}

The only metric compatible with Hubble's findings and the
Cosmological Principle is the Robertson Walker Metric (RWM) with
the corresponding line element,

\begin{equation}
\label{line element} ds^{2}=c^{2}dt^{2}-a^{2}(t) \left[
\frac{dr^{2}}{1-kr^{2}}+r^{2}d\theta^{2}+r^{2} \sin^{2}{\theta}
d\phi^{2} \right].
\end{equation}

For a brief explanation consider the following:
\begin{itemize}
\item For the metric to be homogeneous, isotropic, and obey the
Weyl postulate, the metric must be the same in all directions and
locations, $$g_{\mu \nu}=g_{\mu \mu}.$$
\item For a uniform expansion we must have a scale factor $a(t)$ that is a function
of time only.
\item Allowance for any type of geometry (curvature) must be made.
This is represented by the constant $k$, where $k=0, k=1,$ and
$k=-1$ corresponds to flat, spherical, and hyperbolic geometries,
respectively.
\end{itemize}

There are a few subtleties that must be discussed.  First, the $r$
that appears in the line element (\ref{line element}) is {\bf not}
the radius of the universe.  The $r$ is a dimensionless, comoving
coordinate that ranges from zero to one for $k=1$. The measurable,
physical distance is given by the RWM above.  Choosing a frame
common to two distinct points, one obtains,
$$ds^{2}=c^{2}dt^{2}-a^{2}(t)(1-kr^{2})^{-1}dr^{2},$$ for their
separation.  Where $d\theta$ and $d\phi$ are zero, because one has
freedom to arrange the axis and $ds^{2}$ represents their
separation in spacetime.  Thus, their spatial separation is found
by considering spacelike hypersurfaces, that is $dt^{2}=0$. Thus,
their separation is $$d_{p}=a(t) \int_{0}^{r} (1-kr^{2})^{-1/2} \;
dr.$$  Evidently for a $k=0$ flat universe, the distance is
simply,
\begin{equation} d_{p}=a(t) r.
\label{flatuniverse}
\end{equation} Thus, $a(t)$ has units of length and depends on the geometry of
the spacetime.

The next issue is that of curvature.  The curvature of the
universe is determined by the amount of energy and matter that is
present.  The space is one of constant curvature determined by the
value of $k$. Because any arbitrary scaling of the line element
(\ref{line element}) will not affect the sign of $k$, we have the
following convention\footnote{When the metric is invariant under
multiplication by a scale factor, the metric is said to be {\em
conformally invariant}.}:

\begin{itemize}
\item k=1 represents positive, spherical geometry
\item k=0 represents flat Minkowski space
\item k=-1 represents negative, hyperbolic geometry
\end{itemize}

\subsubsection{The Cosmological redshift}

One observable prediction of an expanding universe is that of
redshifting. When a light wave is traveling from a distant galaxy,
to our own, it must travel through the intervening spacetime. This
results in a stretching of the wavelength of light, since the
spacetime is expanding.  This longer wavelength results in the
light being shifted to a `redder' part of the spectrum.  Of course
light with wavelengths differing from visible light will not be
visible to the human eye, but they will still be shifted to longer
wavelengths.

To quantify this analysis, consider a light ray which must travel
along a null geodesic ($ds^{2}=0$) in the comoving frame with
constant $\theta$ and $\phi$.  Using (\ref{line element}),
$$0=ds^{2}=c^{2}dt^{2}-a^{2}(t)\Big[\frac{dr^{2}}{1-kr^{2}}\Big],$$
so, $$c \: dt=\frac{a(t) dr}{\sqrt{1-kr^{2}}}.$$ Integrating
yields, \begin{equation} \label{cosmicdist}
\int_{t_{e}}^{t_{0}}\frac{c \:
dt}{a(t)}=\int_{0}^{r_{e}}\frac{dr}{\sqrt{1-kr^{2}}}\equiv
f(r_{e}),
\end{equation} where $t_{e}$ is the time the light pulse was emitted,
$t_{0}$ was the time the light pulse was received, and $r_{e}$ is
the distance to the galaxy. Thus, if one knows $a(t)$ and $k$, one
can find the relation between the distance and the time. However,
consider emitting successive wave crests in such a brief time that
$a(t)$ is not given a chance to increase by a significant amount;
i.e., the waves are sent out at times $t_{e}$ and $t_{e}+\Delta
t_{e}$ and received at times $t_{0}$ and $t_{0}+\Delta t_{0}$,
respectively. Then (\ref{cosmicdist}) becomes,

$$\int_{t_{e}+\Delta t_{0}}^{t_{0}+\Delta t_{e}}\frac{c \:
dt}{a(t)}=\int_{0}^{r_{e}}\frac{dr}{\sqrt{1-kr^{2}}}$$

Subtracting (\ref{cosmicdist}) from this equation and using the
fact $a(t)$ doesn't change, one can use the fundamental theorem of
calculus to obtain,

$$\frac{c \: \Delta t_{0}}{a(t_{0})}-\frac{c \: \Delta
t_{e}}{a(t_{e})}=0,$$ or $$\frac{c \: \Delta t_{0}}{c \: \Delta
t_{e}}=\frac{a(t_{0})}{a(t_{e})}.$$ $c \: \Delta t$ is just the
wavelength, $\lambda$. Thus, it follows that the red shift, $z$,
can be defined by \begin{equation} \label{redshift1}
z=\frac{a(t_{0})}{a(t_{e})}-1=\frac{\lambda_{0}}{\lambda_{e}}-1=\frac{\Delta
\lambda}{\lambda_{e}}.
\end{equation} Here $a(t_{0})$ is the scale factor of the universe as measured by
a comoving observer when the light is received, $a(t_{e})$ is the
scale factor when the light was emitted in the comoving frame,
$\lambda_{0}$ is the wavelength observed and $\lambda_{e}$ is the
wavelength when emitted.  It is clear that $z$ will be positive,
since $a(t_{0}) > a(t_{e})$, that is the universe is getting
larger.

In addition to this cosmological redshift, which is due to the
expanding universe, there can also be gravitational redshifts and
Doppler redshifts.  At great distances the former two can be
neglected, but in local cases all three must be considered.

It must also be stressed that the Special Relativity (SR) formula
for redshift can not be used.  This is because SR only holds for
`local' physics.  Attempting to use this across large distances
can result in a contradiction.  For example, the expansion rate of
the universe can actually exceed the speed of light at great
distances.  This is not a violation of SR, because a `chain' of
comoving particles (galaxies) can be put together, spaced so the
laws of SR are not violated.  By summing together the measurements
of each set of galaxies, one finds the expansion rate to exceed
that of light, although locally SR holds locally
\citep{bergstrom}. Another explanation is that in a universe
described by SR, no matter or energy exists and the metric never
changes.  On the contrary, in an expanding spacetime none of these
requirements are true.  Although, SR continues to hold locally,
since a `small enough' region can always be chosen where the
metric is approximately flat.\footnote{One must be careful by what
is meant by `small enough'.  This technical point need not concern
us with the present discussion, see \citep{will}.}

\subsection{The Friedmann Models}

To describe the expansion of the universe one must use the RWM
along with the Einstein equations,
\begin{equation}
\label{einstein equations} R_{\mu \nu}-\frac{1}{2}g_{\mu
\nu}R+\frac{\Lambda g_{\mu \nu}}{c^{2}}=-\frac{8 \pi
G}{c^{4}}T_{\mu \nu},
\end{equation}
to determine the equations of motion.  For reference, a summary of
the metric coefficients, the Christoffel Symbols, and the Ricci
Tensor components are presented in \citep[Chapter 15]{collins}.
Note that in this book the scale factor $a(t)$ is written $R(t)$.

Before proceeding any further, an appropriate stress-energy tensor
must be provided.  This is the difficult part of the process.  The
composition of the known universe is a very controversial topic.
The standard procedure is to consider simplified distributions of
mass and energy to get an approximate model for how the universe
evolves.

At this point, units are chosen such that the speed of light, $c$
is set equal to unity.  This gives the simplification that the
energy density, $\epsilon$ is equal to the mass density, $\rho$
using, $\epsilon=\rho c^{2}=\rho.$  This also allows mass and
energy to be considered together, which is in the spirit of the
stress-energy tensor. The mass/energy density will be referred to
as the energy density for the remainder of this paper. The
stress-energy tensor may be given as:
\begin{equation}
\label{tmunu} T_{\mu \nu}=(p+\rho)u_{\mu}u_{\nu}-pg_{\mu \nu},
\end{equation}
where $p$ is the pressure, $\rho$ is the density, and $u_{\mu}$ is
the four-velocity.

At the earliest epoch of the universe, the contribution of photons
to the energy density would have been appreciable. However, as the
universe cooled below a critical temperature, allowing the photons
to decouple from baryonic matter, the photon contribution became
negligible.  Thus, it is easier to consider different energy
distributions for different epochs in the universe.  The massive
contribution to the energy density is usually referred to as the
Baryonic contribution, since baryons (protons, neutrons, etc.) are
significantly more massive than leptons (electrons, positrons,
etc.) and leptons can therefore be disregarded as a major
contributing factor to the total energy density. There is also the
contribution of vacuum energy, which enters the Einstein equations
through the cosmological constant, $\Lambda$.

For each type of contribution, there is a corresponding density,
$\rho$. The total density can be expressed as the sum of the
different contributions as
\begin{equation}
\rho=\rho_{M}+\rho_{R}+\rho_{\Lambda}.
\end{equation} Furthermore, assuming that one is dealing with a
homogeneous and isotropic fluid, the density can be related to the
pressure by a simple equation of state (see Table \ref{table1}),
\begin{equation}
\label{equation of state} p=\alpha \rho.
\end{equation} Another useful relation involves the Conservation of Energy
(1st Law of Thermodynamics).  Assuming the ideal fluid expands
adiabatically, one finds \citep{finkelstein}, $$dE=-p \: dV,$$
which may be rewritten as,
\begin{equation} \label{conservation eq2} d(\rho
a^{3})=-pd(a^{3}).
\end{equation} Relating (\ref{equation of state}) and (\ref{conservation
eq2}) gives, $$d(\rho a^{3})=-\alpha \rho d(a^{3}).$$  Using the
product rule, $$\rho d(a^{3})+a^{3}d\rho=-\alpha \rho d(a^{3}).$$
Which can be integrated, $$\int \rho^{-1} d\rho = -(1+\alpha) \int
a^{-3} d(a^{3}).$$
\begin{equation}
\label{density to scale factor} \rho \propto a^{-3(1+\alpha)}.
\end{equation}

In the Radiation epoch, where the energy density due to photons
was appreciable (from about $t$=0 to approximately 300,000 years
after the Big-Bang \citep{dekel}), the density due to massive
particles can be neglected.  The pressure is found to be equal to
a third of the density, and we have a value of one-third for
$\alpha$, so $\rho_{R} \sim a^{-4}$ \citep{bergstrom}.

Following this epoch, the Matter Dominated epoch can be modeled
after a `dust' that uniformly fills space. Because the temperature
of the universe had fallen to around 3000 K, most of the particles
had non-relativistic velocities ($v \ll c$). This corresponds to a
negligible pressure and $\alpha$ is therefore zero,
$\rho_{\Lambda} \sim \mbox{constant}$ \citep{bergstrom}.

The last case to consider is that of the vacuum energy.  If the
cosmological constant is indeed nonzero, this form of energy
density will dominate.  For this relation, the pressure is
commensurate with that of a negative density. This would imply a
value of -1 for $\alpha$, so $\rho_{M} \sim a^{-3}$
\citep{bergstrom}. These results are summarized in Table
\ref{table1}.

Given an expression for the energy-momentum tensor, one can now
proceed to find the equations of motion.  The metric coefficients
follow from the Robertson Walker line element, which is given by
equation (\ref{line element}).  Using these coefficients one can
obtain the expression for the left side of the Einstein equations
(\ref{einstein equations}).  Thus, from the Einstein equations one
derives the Friedmann equations in their most general form:
\begin{equation}\label{14} \frac{\ddot{a}}{a}=-\frac{4 \pi G}{3}
\Big( \rho + 3p \Big)+\frac{\Lambda}{3},
\end{equation}
\begin{equation}\label{15}
\Big( \frac{\dot{a}}{a}\Big)^{2}=\frac{8 \pi G
\rho}{3}-\frac{k}{a^{2}}+\frac{\Lambda}{3}.
\end{equation}

Apparently, if the universe is in a vacuum dominated state
$p=-\rho$, (\ref{14}) indicates the universe will be accelerating.
This important conclusion will be the most general requirement for
an inflationary model.

Now is the time to introduce a bit of machinery to make our
calculations more tractable.  Recall that the Hubble constant,
$H(t)$, is defined as \begin{equation}
H(t)=\frac{{\dot{a}(t)}}{a(t)} \label{H(t)}.
\end{equation}
Next, one defines the Deceleration parameter (named for historical
reasons) as,
\begin{equation}
q(t)=-\frac{\ddot{a}(t)}{a(t) H^{2}(t)}. \label{q(t)}
\end{equation} To realize how this term arises, consider the Taylor
expansion of the scale factor, about the present time, $t_{0}$,
\begin{displaymath}
a(t)=
a_{0}+\dot{a}_{0}(t-t_{0})+\frac{1}{2}\ddot{a}_{0}(t-t_{0})^{2}+\ldots,
\end{displaymath} where the sub-zeros indicate the terms are evaluated at the
present. Using equations (\ref{H(t)}) and (\ref{q(t)}), this
becomes
\begin{equation}
a(t) = a_{0} \big[
1+H_{0}(t-t_{0})-\frac{1}{2}q_{0}H_{0}^{2}(t-t_{0})^{2}+ \ldots
\big]. \label{a(t)1}
\end{equation}
Remembering that the crux for obtaining Hubble's Law (\ref{hubble
law}) was measuring the luminosity distance, it is of interest to
consider this calculation quantitatively.  The flux $F$ (energy
per time per area received by the detector) is defined in terms of
the known luminosity $L$ (energy per time emitted in the star's
rest frame) and the luminosity distance $d_{L}$.
\begin{equation}
\label{flux}  F=\frac{L}{4\pi d^{2}_{L}}.
\end{equation} The luminosity distance must take into account the
expanding universe and can be written in terms of the redshift,
$z$ as \citep{turner},
\begin{equation}
\label{lumdist} d^{2}_{L}=a^{2}_{0} r^{2}(1+z)^{2},
\end{equation} where $a_{0}$ is the present scale factor and $r$ is the comoving
coordinate that parameterizes the space. Hubble used the measured
flux and the known luminosity to find the distance to the objects
he measured.  The distance can then be compared with the known
redshift of the object using (\ref{lumdist}) and the velocity can
be approximated.  However, $r$ in (\ref{lumdist}) is not a
observable and it is of interest to examine the great amount of
estimation that must be used to derive the desired result
analytically.

Dividing (\ref{a(t)1}) by $a_{0}$ and making use of
(\ref{redshift1}) yields,
\begin{equation}\label{123}
\frac{a_{0}}{a}=z=H_{0}(t_{0}-t)+\Big( 1+\frac{q_{0}}{2} \Big)
H^{2}_{0}(t_{0}-t)^{2}+ \ldots,
\end{equation} which can be solved for $(t_{0}-t)$,
\begin{equation}\label{345}
(t_{0}-t)=H_{0}^{-1} \Bigg[ z- \Big( 1+\frac{q_{0}}{2} \Big)z^{2}
+ \ldots \Bigg].
\end{equation}
One can also expand (\ref{cosmicdist}) in a power series,
\begin{equation}\label{23}
f(r)=sinn^{-1}(r),
\end{equation} where $r_{e}$ has been replaced by $r$ (for
simplicity) and $sinn^{-1}(r)$ is defined as $sin^{1}(r)$ for
$k=1$, $sinh^{-1}(r)$ for $k=-1$ and $r$ for $k=0$. So to lowest
order, (\ref{cosmicdist}) can be estimated as $r$, and the l.h.s.
of (\ref{cosmicdist}) can be estimated as,
\begin{equation}
\int^{t_{0}}_{t_{e}} \frac{c \; dt}{a(t)} \approx
\frac{c(t-t_{0})}{a_{0}}.
\end{equation}  Using the approximation from (\ref{23}) and the
above result we have, $$\frac{c(t-t_{0})}{a_{0}} \approx r.$$
Substitution of $t-t_{0}$ from (\ref{345}) and keeping only lowest
order terms yields, $$a_{0}r \approx c(t-t_{0}) \approx
\frac{cz}{H_{0}}.$$  At small redshift, $z \ll 1$ one finds $z
\approx v/c$.  Thus, making this final approximation one obtains,
\begin{equation}
cz \approx v \approx a_{0}H_{0}r \approx H_{0} d_{p},
\end{equation}
\begin{equation}
v \approx H_{0} d_{p},
\end{equation} where $d_{p}$ is the physical distance.  Thus, we have obtained
Hubble's law (\ref{hubble law}) as an approximation. This
derivation reflects the reason that the law only holds locally.
The number of approximations that were needed to proceed was
appreciable. Furthermore, one finds that this law deviates
significantly at large $z$ as one would expect.

For a matter dominated model, one finds the exact Hubble relation
to be given by \citep{turner},
\begin{equation}
H_{0} d_{l} = q_{0}^{-2}
\Bigg[zq_{0}+(q_{0}-1)\Big(\sqrt{2q_{0}z+1}-1\Big) \Bigg],
\end{equation} which depends on the deceleration parameter,
$q_{0}$, which in turn relies on the curvature and the total mass
density of the universe.

\subsection{Matter Dominated Models} The present epoch is
best described by a matter dominated universe, so it is perhaps
best to explore this model first. Again, matter domination
corresponds to a non-relativistic, homogeneous, isotropic `dust'
filled universe with zero pressure. By setting $p=0$ in (\ref{15})
and incorporating the $\Lambda$ term into a total density,
$\rho_{T}=\rho+\frac{\Lambda}{8 \pi G}$, the Friedmann equations
for a matter dominated universe emerge,
\begin{equation}
\Big( \frac{\dot{a}}{a}\Big)^{2}+\frac{k}{a^{2}}=\frac{8 \pi G
\rho_{T}}{3}. \label{friedmann2}
\end{equation}
\begin{equation}
2 \frac{\ddot{a}}{a}+\Big(
\frac{\dot{a}}{a}\Big)^{2}+\frac{k}{a^2}=0. \label{friedmann1}
\end{equation} Again, the value of $k$ describes the geometry of the
space. Equation (\ref{friedmann1}) is actually obtained by
combining (\ref{14}) and (\ref{15}) and is often called the
acceleration equation.

The idea of the {\em total} density, $\rho_{T}$, might be a bit
confusing since it has been stated that the model is matter
dominated. Although this is true, there can still be a small
contribution in the form of radiation and other forms of energy,
such as dark matter. The point is that any of these should be much
less than $\rho_{M}$ for the model to be accurate. It will also be
seen that $\rho_{M}$ can also be broken into different
contributions as tacitly stated in the previous remark about dark
matter.  For the remainder of this section we take $\rho$ to mean
$\rho_{T}$ to keep the notation as simple as possible.

\subsubsection{The Einstein-DeSitter Model}

The Einstein-DeSitter model is a matter dominated Friedmann model
with zero curvature ($k=0$).  This model corresponds to a
Minkowski universe (zero curvature), in which the universe will
continue to expand forever with just the right amount of energy to
escape to infinity.  It is analogous to launching a rocket.  If
the rocket is given insufficient energy, it will be pulled back by
the Earth. However, if its energy exceeds a certain critical
velocity (escape velocity), it will continue into space with ever
increasing speed. If it has exactly the escape velocity, it will
proceed to escape the Earth with a velocity going to zero as the
rocket approaches spatial infinity. The Einstein-DeSitter model
corresponds to the universe having exactly the right escape
velocity provided by the Big-Bang to escape the pull of gravity
due to the matter in the universe.

By substituting $k=0$ into (\ref{friedmann1}) and
(\ref{friedmann2}), the Friedmann equations become
\begin{equation}
0=2\frac{\ddot{a}}{a}+\Bigg( \frac{\dot{a}}{a} \Bigg)^{2}=-2 q(t)
H^{2}(t)+H^{2}(t), \label{k0friedmann2}
\end{equation}
\begin{equation}
\Bigg( \frac{\dot{a}}{a} \Bigg)^{2}=H(t)^{2}=\frac{8 \pi G
\rho}{3}. \label{k0friedmann1}
\end{equation} By solving (\ref{k0friedmann1}) for $\rho$, a critical
density can be found for a flat universe. The critical density is
the amount of matter required for the universe to be exactly flat
($k=0$) and is a function of time.  The critical density at the
present is defined as,
\begin{equation}
\label{critical density} \rho_{c}=\frac{3H^{2}_{0}}{8 \pi G}.
\end{equation}
If the density of the universe exceeds the critical density, the
universe is open.  Conversely, if the density is below $\rho_{c}$
the universe is open.  For the observed Hubble parameter as
defined in (\ref{hubble const}), the critical density today
corresponds to a value,
\begin{equation}
\rho_{c} \approx 2 \: h_{0}^2 \times 10^{-23}
\frac{\mbox{g}}{\mbox{m}^{3}}.
\end{equation} This is equivalent to roughly 10 hydrogen atoms per cubic meter.
Although, this is incredibly small compared to Earthly standards,
it must be remembered that most of space is empty and the concern
is the total energy density.

Notice that the critical density depends on the Hubble constant.
This means that the density required for a flat universe will
change with time, in general, as the universe expands.  For the
universe to be `fine-tuned' to this precision is highly
improbable; yet, most observations suggest this type of geometry.
This paradoxical issue is referred to as the Flatness problem and
will lead to one of the claimed triumphs of Inflation theory.
Because it is believed that the universe is so close to being
flat, it is useful to define the density parameter, $\Omega$.
$\Omega_{0}$ is the ratio of the density observed today,
$\rho_{0}$, to the critical density, $\rho_{c}$. In general,
$\Omega$ is the ratio of the density to the critical value.

The quantity $\Omega$ together with Equation (\ref{k0friedmann2}),
which implies $q_{0}=\frac{1}{2}$, can be used to discriminate
between the possible geometries for the matter dominated universe
(see Table \ref{table2}).

From the previous result for a matter dominated energy density, we
found $\rho \sim a^{-3}$.  From this relation, conservation of
energy follows,  $$\frac{d}{dt}(\rho a^{3})=0 \Rightarrow \rho
a^{3}= \mbox{constant}.$$ This can be used to obtain a useful
relation for $\rho$, $$\rho=\rho_{0} \frac{a^{3}_{0}}{a^{3}}.$$
Returning to the Friedmann equation (\ref{k0friedmann1}) and
substituting the above expression for $\rho$ one finds,

$$\Bigg( \frac{\dot{a}}{a} \Bigg)^{2}=\frac{8 \pi G
\rho}{3}=\frac{8 \pi G \rho_{0}}{3}  \frac{a^{3}_{0}}{a^{3}} .$$
Combining terms in $a$, $$a\dot{a}^{2}=\frac{8 \pi G \rho_{0}
a_{0}^{3}}{3},$$ now integrating, $$\int_{0}^{a}
\sqrt{a}da=\int_{0}^{t} \sqrt{\frac{8 \pi G \rho_{0}
a_{0}^{3}}{3}}dt,$$

$$a(t)=a_{0} \Bigg( \frac{8 \pi G \rho_{0}}{3}
\Bigg)^{\frac{1}{3}}t^{\frac{2}{3}},$$

\begin{equation}
a \propto t^{2/3}.
\end{equation} So for the Einstein-DeSitter Model, the scale factor evolves as
$t^{2/3}$.

\subsubsection{The Closed Model} The Closed Model is characterized
by a positive curvature, $k=1$. Thus, the spatial structure is
that of the 3-sphere, similar to the surface of a sphere, but in 3
dimensions instead of 2. This model corresponds to a universe that
begins at a `Big-Bang' and continues to expand until gravity
finally halts the expansion. The universe will then collapse into
a `Big-Crunch', which will resemble the reverse process of the
`Big-Bang'. The ability of the matter (or energy) in the universe
to halt the expansion obviously depends on the density. If the
matter-energy density is too low, the universe will have enough
momentum from the `bang' to escape the pull of gravity.  In the
Closed Model the density of the universe is great enough to halt
the expansion and start a contraction. This corresponds to a value
of $\Omega>1$, which is evident from the use of the Friedmann
equations with $k=1$. Plugging this $k$ value into the Friedmann
equations (\ref{friedmann1}),(\ref{friedmann2}) and using
$\ddot{a}=-qH^{2}a$ one gets,
$$2\frac{\ddot{a}}{a}+\frac{{\dot{a}}^{2}}{a^2}+\frac{1}{a^2}=2
\Big(-qH^{2} \Big) + H^{2} + \frac{1}{a^{2}}=0.$$ This can be
expressed as \begin{equation} \label{k1friedmann1}
\frac{1}{a^{2}}=H^{2} \Big[2q-1 \Big].
\end{equation}
Equation (\ref{friedmann2}) takes the form,
\begin{equation}
\label{k1friedmann2}
\frac{{\dot{a}}^{2}}{a^{2}}+\frac{1}{a^{2}}=\frac{8 \pi G
\rho}{3}.
\end{equation} Combining (\ref{k1friedmann1}) and (\ref{k1friedmann2}) gives,
$$H^{2}+ \Bigg[ H^{2} \Big( 2q-1 \Big) \Bigg] = \frac{8 \pi G
\rho}{3},$$ or $$2 q H^{2} = \frac{8 \pi G \rho}{3}.$$ Thus,
\begin{equation}
\label{rhoclosed} \rho = \frac{3 H^{2} q}{4 \pi G}.
\end{equation}

Comparing (\ref{rhoclosed}) with the critical density
(\ref{critical density}) and the value of $q>1/2$ in Table
\ref{table2}, it is evident that $\rho>\rho_{c}$ for the universe
to be closed.  In terms of $\Omega$, this gives
\begin{equation}
\Omega = \frac{\rho}{\rho_{c}} > 1.
\end{equation}

The advantage of equation (\ref{rhoclosed}) above, is that the
density is expressed all in quantities that can be measured. In
that, if 2 of the 3 quantities are known the third may be found.

\subsubsection{The Open Model} The so-called Open Model\footnote{Although it is a standard practice to refer to this case
$(k=-1)$ as the `Open' Model, it should be noted that the model
can actually correspond to a closed universe. This is the result
of a non-trivial topology, which results in geometry that can be
hyperbolic; but, the topology can cause it to be contained in a
finite space \citep{cornish,luminet}.} is the case where $k=-1$
and the geometry is said to be hyperbolic. Taking $k=-1$ in the
Friedmann equations (\ref{friedmann1}),(\ref{friedmann2}),

\begin{equation}
\label{k-1friedmann1} 2
\frac{\ddot{a}}{a}+\frac{{\dot{a}}^{2}}{a^2}-\frac{1}{a^2}=0.
\end{equation}

\begin{equation}
\label{k-1friedmann2} \frac{\dot{a}^{2}}{a^{2}}-\frac{1}{a^{2}}=
\frac{8 \pi G \rho}{3}
\end{equation}

Solving these equations
(\ref{k-1friedmann1}),(\ref{k-1friedmann2}) yields the same value
of the density (\ref{rhoclosed}) as the closed model, but in this
case $q< 1/2$ and $\Omega<1$, as previously discussed.

\subsection{Summary}

\begin{itemize}
\item All cosmological models are characterized by `test particles',
which are galaxies that are distributed in a homogeneous and
isotropic manner in accordance with the CP.
\item {\em Open Models} are characterized by $\Omega<1$, negative
curvature ($k=-1$), hyperbolic geometry, a deceleration parameter
$q<1/2$, and infinite spatial extent (ignoring topology).
\item {\em Closed Models} are characterized by $\Omega>1$, positive curvature
($k=1$), spherical geometry, a deceleration parameter $q>1/2$ and
finite spatial extent.
\item {\em Flat Models} are characterized by $\Omega=1$, flat geometry with
no curvature ($k=0$), infinite spatial extent, a deceleration
parameter $q=1/2$ and with an age corresponding to
Age=$H_{0}^{-1}$, since the Hubble Constant is in-fact constant.
\end{itemize}
%------------------------------------End Section One -------------------------
\newpage
%------------------------------------Begin Section Two------------------------
\section{A Brief History of the Universe}

One of the successes of the hot Big Bang model is its prediction
of the light elements.  These predictions are verified by
observations of the structure and composition of the oldest stars,
quasars, and other quasi-stellar remnants (e.g., QSOs)
\citep{9907128,9904407,9712031,9904223}. The process by which the
elements form is referred to as nucleosynthesis.

The Hot Big-Bang model predicts a universe that will go through
several stages of thermal evolution. As the universe expands
adiabatically, the temperature cools, scaling as $$T \propto
a^{-3(\gamma-1)}.$$ Here $\gamma$ is the ratio of specific heats
and is equal to $4/3$ for a radiation dominated universe
\citep{peacock}. This relation is manifest, since the temperature
is equivalent to the energy density divided by the volume (in
natural units $\hbar=c=k=1$).  Moreover, the radiation energy
density is redshifted by an additional factor of $1/a(t)$ since
the Hubble expansion stretches the wavelength, which is inversely
proportional to the energy: $$\rho_{R}\propto a^{-4}, $$ and
$$\rho_{R}=a^{-3}E.$$ Setting the Boltzman constant to unity,
\begin{equation}
E=T \propto a^{-1}. \label{temp}
\end{equation} This can also be understood using the
DeBroglie wavelength of the photon (for radiation)
\citep{peebles}. The wavelength is inversely proportional to the
energy in natural units.  This raises the issue of a possible
factor of redshifting for the DeBroglie wavelength of a massive
particle.  For a particle the simple relation $E=pc$ does not
hold;  thus, the velocity of the particle can decrease to preserve
its wavelength.  This redshifting is analogous to that of equation
(\ref{peculiar velocity}) in the first section and gives an
alternative explanation of particles in motion settling into the
cosmic rest frame.  This also explains why one might expect to
find primarily non-relativistic (cold) matter, which just means
particles traveling at speeds much less than $c$.

From relation (\ref{temp}) for the temperature, one has a
quantitative way to find critical temperature scales in the
evolution of the universe. For a given value of the curvature the
relation between the scale factor and time can yield an expression
between temperature and time.  For example, in a matter dominated,
flat universe,

$$ T \propto t^{-2/3},$$ which was derived earlier.

In thermal physics one is usually interested in thermal
equilibrium. This consideration is accounted for by the condition,
$$\frac{\Gamma}{H} \gg 1.$$ This relations shows that the reaction
rate, $\Gamma$, must be much greater than the rate at which the
universe expands for thermal equilibrium to be reached. $\Gamma$
is related to the cross-section of the given particle interaction
by $$\Gamma=n\sigma |v|,$$ where $\sigma$ is the cross-section,
$|v|$ is the relative speed, and $n$ is the number density of the
species\footnote{For the reader interested in learning more about
particle interactions see, \citep[Chapter
~7]{bergstrom},\citep[Chapter~9]{peacock}}.

From equation (\ref{temp}), one can see that the universe began as
a point of infinite temperature and zero size.  This is a singular
point for the history of the universe and the standard Big Bang
model (General Relativity) breaks down at this
singularity\footnote{Supersting Theory offers solutions to the
problem of a singularity by setting an ultimate smallness, the
Planck length $(10^{-33}cm)$.  This is outside the scope of this
paper, but the reader is referred to \citep{greene} for an
excellent popular account of strings and cosmology and in
\citep{gasperini} there are a number of papers with rigorous
treatments of string cosmology.}. However, after the Planck time
$(10^{-43})$s one can follow the evolution using the concepts of
thermal physics and particle theory\footnote{For an introductory
survey of particle theory see \citep{moyer,tipler} and for
particle physics in cosmology \cite{rolfs} is an excellent book.
The remainder of this paper will assume a basic knowledge of
particle theory and thermal physics.}.

In the earliest times following the Planck epoch, all matter
existed as free quarks and leptons. The existence of free quarks
(known as asymptotic freedom) is made possible by the high energy
(temperature) during the early moments of the Big Bang. The
universe cools, as indicated by (\ref{temp}), and  the quarks
begin to combine under the action of the strong force to form
nucleons. This phase of formation is referred to as Baryogenesis,
because the baryons (e.g., protons and neutrons) are created for
the first time. The expansion continues to allow leptons, such as
electrons, to interact with nucleons to form atoms.  At this
point, referred to as recombination, the photons in the universe
are free to travel with virtually no interactions. For example,
hydrogen is the most abundant element to form in nucleosynthesis
and at the time of decoupling the temperature of the photons has
dropped to around $3000\;K$.  This corresponds to less than
$13.6\; \mbox{eV}$, the energy needed to ionize the atoms.
Therefore, there are no longer free electrons to interact with the
photons and in fact the energy of a photon (around $.26 \;
\mbox{eV}$, at this temperature) is so low that it can not
interact with the atoms. In this way the photons have effectively
decoupled from matter and travel through the universe as the
cosmic background discussed in Part II. A brief summary of the
most significant events are encapsulated below,

\begin{itemize}
\item $10^{-4}$ seconds: Baryogenesis occurs, quarks condense under strong interaction
to form nucleons (e.g., Protons and Neutrons)

\item 1 second: Nucleosynthesis occurs, universe cools enough (photon energies $\sim \;1 \;\mbox{MeV}$) for light nuclei to form
(e.g., deuterons, alpha particles).

\item $10^{4}$ years: Radiation density becomes equal to matter
density, since the radiation density has extra factor of $a^{-1}$
due to red-shifting.  Matter density is the dominate energy
density after this epoch.

\item $10^{5}$ years: Recombination occurs and electrons are
combined with nucleons to form atoms.  This time also coincides
with the decoupling of photons from matter, giving rise to a
surface of last scattering of the cosmic background radiation.

\item $10^{10}$ years: The present.
\end{itemize}
%------------------------------------End Section Two -------------------------
\newpage
%------------------------------------Begin Section Three----------------------
\section{Problems with the Standard Cosmology}

The hot Big Bang model has been very successful in predicting much
of the phenomena observed in the universe today.  The model
successfully accounts for nucleosynthesis and the relative
abundance of the light elements, (e.g., Hydrogen $\sim$~75\%,
Helium $\sim 25\%$, Lithium (trace), Berylium (trace)).  The
prediction of the Cosmic Background Radiation and the fact that
the universe is expanding (i.e., The Hubble Law), both represent
successful predictions of the Big Bang theory. However, this model
suggests questions which it can not answer, which brings about its
own demise. These anomalies are discussed in the following
sections.

\subsection{The Horizon Problem}  Why is the universe so
homogeneous and isotropic on large scales?  Radiation on opposite
sides of the observable universe today appear uniform in
temperature.  Yet, there was not enough time in the past for the
photons to communicate their temperature to the opposing sides of
the visible universe (i.e., establish thermal equilibrium).
Consider the comoving radius of the causally connected parts of
the universe at the time of recombination compared to the comoving
radius at the present, found from Equation (\ref{cosmicdist})
(remember $c=\hbar=1$).
\begin{equation}\label{horizonproblem}
\int_{0}^{t_{rec}} \frac{dt}{a(t)} \; \ll \int_{t_{rec}}^{t_{0}}
\frac{dt}{a(t)}.
\end{equation}
This means a much larger portion of the universe is visible today,
than was visible at recombination when the CBR was `released'. So
the paradox is how the CBR became homogeneous to 1 part in
$10^{5}$ as we discussed in Part II.  There was no time for
thermal equilibrium to be reached.  In fact, any region separated
by more than 2 degrees in the sky today would have been causally
disconnected at the time of decoupling \citep{liddleeprint}.

This argument can be made a bit more quantitative by consideration
of the entropy, $S$, which indicates the number of states within
the model.  This can be used as a measure of the size of the
particle horizon \citep{turner}.
\begin{equation}
\label{entropyrd} S^{RD}_{Horizon}=s\frac{4}{3}\pi t^{3} \approx
0.05 \: g_{*}^{-1/2}(m_{pl}/T)^{3},
\end{equation}
\begin{equation}
\label{entropymd} S^{MD}_{Horizon}=s\frac{4}{3}\pi t^{3} \approx 3
\times 10^{87} (\Omega_{0}h^{2})^{-3/2}(1+z)^{-3/2},
\end{equation}
where $m_{pl}$ is the Planck mass, $s$ is the entropy density,
$g_{*}$ is the particle degeneracy, and $z$ is the redshift. These
equations for the entropy of the horizon in a radiation dominated
(\ref{entropyrd}) and matter dominated universe (\ref{entropymd}),
are presented only to motivate the following estimates. For an
explanation please consult \citep{turner}.

At the time of recombination ($z \approx 1100$), when the universe
was matter dominated, equation (\ref{entropymd}) gives a value of
about $10^{83}$ states.  Compared with a value today of $10^{88}$
states, this is different by a factor of $10^{5}$.  Thus, there
are approximately $10^{5}$ causally disconnected regions to be
accounted for in the observable universe today.  The hot Big Bang
offers no resolution for this paradox, especially since it is
assumed to be an adiabatic (constant entropy) expansion.

\subsection{The Problem of Large-Scale Structure}
In contrast to the horizon problem, the fact that the Big Bang
predicts no inhomogeneity is a problem as well.  How are galactic
structures to form in a perfectly homogeneous universe? The fact
that galaxies have been shown to cluster locally with great voids
on the order of 100 Mpc, is proof of the inhomogeneity of the
universe.  Moreover, the $10^{-5}$ anisotropies (temperature
differences) on angular scales of 10 degrees as measured by the
COBE satellite, form a blueprint of the seeds of formation at the
time of decoupling. However, there is no mechanism within the Big
Bang theory to account for these `seeds', or perturbations, that
result in the large-scale structure. Not only does the Big Bang
predict homogeneous structure, but it also had to `explode' in
just the right way to avoid collapse. This is often called the
fine-tuning problem. Cosmologists would like to have a theory that
does not require specific parameters to be put in the theory ad
hoc. The density, the expansion rate, and the like, prove to be
other unfavorable aspects of the hot Big Bang.

\subsection{The Flatness Problem}
The flatness problem is another example of a fine-tuning problem.
The contribution to the critical density by the baryon density,
based on calculations from nucleosynthesis and the observed
abundance of light elements, are in good agreement with
observations and give $\Omega_{B} < 0.1$.  The radiation density
is negligible and it is believed that non-baryonic dark matter, or
quintessence/dark energy (non zero cosmological constant), will
contribute the remainder of the critical density, yielding
$\Omega=1$. Although, an $\Omega$ anywhere within the range of 1
causes a problem.

The Friedmann equation (\ref{friedmann2}) can be used to take into
account how $\Omega$ changes with time. Noting that $H=\dot{a}/a$
and $\Omega=\rho / \rho_{c}$, one can divide (\ref{friedmann2}) by
$H^{2}$ to obtain, $$\mid \Omega(t)-1 \mid=\frac{\mid k
\mid}{a^{2}H(t)^{2}}.$$ Using the relationships between the scale
factor and time, $$\mbox{Matter Domination:} \;\;\;\; a_{M} \sim
t^{2/3},$$ $$\mbox{Radiation Domination:} \;\;\;\; a_{R} \sim
t^{1/2},$$ and using the definition of $H$ yields, $$\mbox{Matter
Domination:} \;\;\;\;\mid \Omega(t)-1 \mid \sim t^{2/3},$$
$$\mbox{Radiation Domination:} \;\;\;\; \mid \Omega(t)-1 \mid \sim
t,$$ From these relations one can see that $\Omega$ must be very
fine-tuned at early times.  For example, requiring $\Omega$ to be
one today, corresponds to a value of $\mid \Omega(1)-1 \mid \sim
10^{-16}$ at the time of decoupling and a value of $\mid
\Omega(10^{-43})-1 \mid \sim 10^{-60}$ at the Planck epoch. This
value seems unnecessarily contrived and indicates that we live at
a very special time in the universe.  That is to say, when the
universe happens to be flat.  An alternative is that the universe
has been, is, and always will be flat.  However, this is a very
special case and it would be nice to have a mechanism that
explains why the universe is flat.  The Big Bang offers no such
explanation.

\subsection{The Monopole Problem}
At early times in the expansion ($z>1000$), the physics of the
universe is described by particle theory.  Many of these theories
predict the creation of topological defects.  These defects arise
when phase transitions occur in particle models.  Since the
temperature of the universe cools as the expansion proceeds, these
phase transitions are natural consequences of symmetry breakings
that occur in particle models. Several types of defects are
described briefly below \citep[Chapter~10]{peacock},

\begin{itemize}
\item {\em Domain Walls} -- Space divides into connected regions; one
region with one phase and the other region exhibiting the other
phase.  The regions are separated by walls of discontinuity
described by a certain energy per unit area.
\item {\em Strings} -- These are linear defects, characterized by
some mass per unit length.  They can be visualized at the present
time as large strings stretched in space that possibly cause
galaxies to form into groups.  They serve as an alternative to
inflation, for explaining the large-scale structure of the
universe.  However, at the moment they are not favored due to lack
of observations of the gravitational-lensing effect they should
exhibit\footnote{Also note that these are not visible objects,
they are distortions in the space-time fabric.}.

\item {\em Monopoles} -- These are point defects, where the field
points radially away from the defect, which has a characteristic
mass.  These defects have a magnetic field configuration at
infinity that makes them analogous to that of the magnetic
monopole, hypothesized by Maxwell and others.
\item {\em Textures}  These objects are hard to visualize and are
not expected to form in most theories.  One can consider them as a
kind of combination of all the other defects.
\end{itemize}

Out of all these defects, monopoles are the most prevalent in
particle theories.  It becomes a problem in the hot Big Bang
model, when one calculates the number of monopoles produced in
events, such as the electroweak symmetry breaking. One finds they
would be the dominate matter in the universe.  This is contrary to
the fact that no monopole has ever been observed, directly or
indirectly, by humans.  These monopoles would effect the curvature
of the universe and in turn the Hubble parameter, galaxy
formation, etc. Therefore, unwanted relics, such as monopoles,
remain an anomalous component of the hot Big Bang theory.
%------------------------------------End Section Three -----------------------
\newpage
%------------------------------------Begin Section Four-----------------------
\section{The Inflationary Paradigm}

In past years, inflation has become more of a scenario than model.
A plethora of models have been suggested, all of which share the
common feature that the universe goes through a brief period of
rapid expansion.  This rapid expansion is manifested in the
evolution of the scale factor, $a(t)$.  In the case of inflation,
$a(t) \sim t^{n}$, where $n>1$ and the universe expands faster
than light. This does not violate relativity, since the spacetime
is the thing expanding (i.e., no information is being
transferred). Since $n$ can take on any value greater than one,
this is already an example of the flexibility of the theory.

In Section II it was shown by Equation (\ref{14}) that if an
equation of state $p=-\rho$ is achieved and one has a positive
cosmological constant, then the universe will accelerate.
Incorporating the cosmological constant, $\Lambda$, into an energy
density, $\rho_{\Lambda}$, and assuming it is the dominate one can
use ($\ref{15}$) and ($\ref{H(t)}$) to obtain,
\begin{equation}
H^{2}=\Big( \frac{\dot{a}}{a} \Big)^{2}=\frac{8\pi G
\rho_{\Lambda}}{3}-\frac{k}{a^{2}}, \label{friedconstant}
\end{equation}
One can choose to ignore the curvature term, since one anticipates
a large increase in the scale factor.  That is, the presence of
the scale factor in the denominator of the $-{k}/{a^{2}}$ term in
the equation above will leave this term negligible. This is often
referred to as the redshifting of the curvature, since the effect
of the curvature can be ignored if the scale factor becomes large
enough during a period of constant energy density
$\rho_{\Lambda}$. This is actually a glimpse of how the flatness
problem will be resolved. So, ignoring the curvature term, we have
\begin{equation}
\Big( \frac{\dot{a}}{a} \Big)^{2}=\frac{8\pi G \rho_{\Lambda}}{3}.
\end{equation}

This is a differential equation with the solution,
\begin{equation}
a(t) \sim e^{Ht} \label{a_exp},
\end{equation}
where $H=\Big( \frac{8}{3} \pi G \rho_{\Lambda} \Big)^{1/2}$ and
since $\rho_{\Lambda}$ is a constant, so is $H$. This model is
referred to as the DeSitter model\footnote{Not to be confused with
the Einstein-DeSitter model.}. By introducing a negative pressure,
the flatness problem is solved. The crux of this argument is that
$\rho_{\Lambda}$ is a constant\footnote{Actually, $\Lambda$ does
not have to be a constant; in-fact, it can be a function of time.
Such vacuum energies are referred to as Quintessence, or Dark
energy, and are the subject of much research.  Unfortunately, time
will not permit a discussion
(\citep{0001051},\citep{9908518},\citep{9912046}).}. This comes
from the fact that $\rho_{\Lambda}$ is an intrinsic property of
the spacetime manifold.  As the manifold is stretched, this vacuum
energy does not change.  Another way this can be explained is by
that the Einstein equations are arbitrary up to a constant term
$\Lambda$. The disadvantage of this explanation is that it does
not manifest the connection between cosmology and particle theory
(more on this later).

Since $\rho_{\Lambda}$ is taken to be the dominate form of energy,
the other contributions to the density in the Friedmann equation
($\ref{friedmann2}$) are also redshifted away, since $\rho_{M}
\sim a^{-3}$, $\rho_{R} \sim a^{-4}$.  This leads to the
conclusion that no matter what the initial distribution of
$\rho_{T}=\rho_{{M}}+\rho_{R}+\rho_{\Lambda}$, the vacuum energy
will eventually dominate.  Thus, the assumption of
$\rho_{\Lambda}$ domination can actually be relaxed.

So given a constant vacuum term, the DeSitter scenario `drives'
the universe to a flat geometry, thus approaching $\rho=\rho_{c}$,
where $\rho_{c}$ is the critical density, (i.e.,
$\rho_{c}={3H^{2}}/{8\pi G}$). This evolution, if allowed to
continue, will produce an empty universe with practically no
radiation or matter.  The fact that we live in a universe that is
full of matter and radiation is why the original proposal, by
DeSitter, was rejected and forgotten.

The revision of this idea was suggested by Guth in the early
1980's \citep{guth},\citep{guth2}.  The crux to the modern
inflationary scenario, in contrast to the DeSitter model, is to
limit the amount of time that this rapid expansion (inflation)
occurs. Guth explained the physical mechanism for such an
inflationary period as corresponding to a phase transition in the
early universe.  By limiting the time of the quasi-exponential
expansion, Guth was able to produce a universe more like our own.
Unlike DeSitter's model, which was based on a pure solution to
Einstein's equations, Guth's idea was based on ideas from particle
physics.  Guth was studying a class of grand unified theories
(GUTs) and the predictions they make about particle production in
the universe. This suggested how cosmology could be united with
particle physics in a phenomenological manner, which has become
one of the most appreciated beauties in modern physics today.

\subsection{Particle Physics and Cosmology}
To better understand the motivation behind inflation, it is
important to outline a few aspects of particle physics.  Often
inflation is introduced in an abstract and unaesthetic manner. One
speaks of an inflaton field, an arbitrary scalar field, for which
there is no physical motivation. This is often the case because
this type of introduction requires limited knowledge of the
relevant topics.  This includes, but is not limited to, the
relativistic Schr{\"o}dinger (Klein-Gordon) equation, the Dirac
equation, scalar fields, symmetries, and group theory.

Since this paper is intended for undergraduates, a brief summary
is presented on how one can pursue this knowledge in a qualitative
and brief manner. A brief overview of the concepts in particle
theory will be provided as needed. Thus, the reader is presented
with a dilemma. One may choose to pause at this point and do a
brief survey of the suggested texts or one may continue and plan
to fill in the details at a later time.  Both options have their
advantages and disadvantages.  I chose the former.

From the author's experience, a student should read through all of
the references to get an intuitive picture of the theory and then
go back and comb through the details and `hairy' calculations.

Three possible routes to obtaining the knowledge needed to
continue are,
\begin{itemize}
\item Thorough Route (The one the author took)\\
\citep[Chapter~15-16]{weinberg3}--Introduction to cosmology with
general relativity

\citep[Chapter~13-14]{tipler},\citep{griffiths2}--Elementary
introduction to particle theory

\citep[Chapter~1-6]{strange}--Introduction to relativistic quantum
mechanics

\citep{kaku}--Introduction to quantum field theory

\citep{peacock}, \citep{bergstrom}--Bring the picture together

\item Fast Route\\
\citep{peacock},\citep{bergstrom}--Bergstr{\"o}m extracts the
particle physics to the appendix, so as not to interfere with the
focus. Both of these books are excellent and I also recommend,
\citep{collins}.

\item Very fast route
\citep[Appendix~B~and~C]{bergstrom}
\end{itemize}

\subsubsection{A Brief Summary of the Modern Particle Physics}

There are four fundamental forces in the realm of physics today;
gravitation, electromagnetism, the weak force, and the strong
force.  For most of the twentieth century, physicists have worked
vigorously to combine or unify these forces into one, in much the
same way Maxwell combined the seemingly disparate forces of
electricity and magnetism.  Great progress has been made to unify
three of the four forces, excluding the realm of gravitation. The
first breakthrough came with the unification of electromagnetism
and the weak force into the electroweak force\footnote{As an aside
to the interested reader, the electroweak force is not really a
unified force, in the strict sense of the world, because the
theory contains two couplings. See \citep{kaku} for more.} . This
work was done primarily by Glashow, Salam, and Weinberg
\citep{salam},\citep{weinberg2} in the late sixties. Although
their theory was not realized until 1971, when the work of 'tHooft
showed their theory and all other Yang-Mills theories could be
renormalized \citep[Chapter~1]{kaku}. Later work was done to unify
the strong and electroweak under the symmetry\footnote{See
\citep{ryder} for a description of symmetries and how they relate
to particle physics.}, $$SU(3) \; \otimes \; SU(2) \; \otimes \;
U(1).$$ This model is referred to as the Standard Model and has
made a number of predictions, which have been verified by
experiment. However, there are many aspects of the model that
suggest it is incomplete. The model produces accurate predictions
for such phenomena as particle scattering and absorption spectra.
Although, the model requires the input of some 19 parameters.
These parameters consist of such properties as particle masses and
charge. But one would hope for a model that could explain most, if
not all of these parameters. This can be accomplished by taking
the symmetry group of the standard model and embedding it in a
higher group with one coupling.  This coupling, once the symmetry
is broken, would result in the parameters of the standard model.
Theories of this type are often referred to as grand unified
theories (GUTs). Many such models have been proposed along with
some very different approaches. Some current efforts go by the
interesting names; Superstring theory, Supersymmetry, Technicolor,
SU(5), etc.

Of all the proposed theories the most promising at the current
moment is Superstring theory.  In addition to unifying the three
forces, this theory can also include the fourth force, gravity.
These theories (there's more than one) can be summarized quite
simply. In the standard model, and in all undergraduate physics
courses, particles are considered points.  If you have ever given
any thought to this, it mostly likely has troubled you.  You are
not alone and the creators of string theory had this very idea as
their motivation.  String theory assumes that particles are not
points, instead they are tiny vibrating strings.  The modes of
vibration of the string give rise to the particle masses, charges,
etc.  This simple picture, along with the idea of supersymmetry,
produces a model that presents the standard model as a low energy
approximation.

Supersymmetric theories differ from the standard model, by the
existence of a supersymmetric partner for each particle in the
standard model. For example, for each half-integer spin lepton
there corresponds an integer spin slepton (thus, it is a boson).
These supersymmetric partners are not observed today, because they
are extremely unstable at low temperatures. However, some versions
of the theory suggest a conservation of supersymmetric number.  If
this is the case, then all of the supersymmetric particles would
be expected to decay into a lowest energy mode referred to as the
neutralino.  As a result, this particle is one of the leading
candidates for cold dark matter \citep[Chapter~6]{bergstrom}.

The link with cosmology is further exhibited because the hot Big
Bang model predicts that at some time in the past, the temperature
was high enough for GUTs to be tested. Because it is impossible to
recreate these temperatures today, the universe offers the only
experimental apparatus to examine the physics of these unified
theories\footnote{This statement is not truly accurate. Particle
theories, such as GUTs, will be further verified with the
detection of the symmetry breaking, or Higgs particle. This
particle should be detectable around 1Tev, which is currently
possible.}. As the universe expands, and thus cools ($T \sim
a^{-1}$), the supersymmetry is broken and the particles manifest
themselves as the different particles that we observe today.

Superstring theorists have attempted to unify these supersymmetric
models with gravity into a so-called Theory Of Everything (TOE).
Some theories have relaxed the supersymmetric requirement and
still produce TOEs by the addition of higher dimensions. Some
proposed TOEs worth mentioning are: Superstrings, M-Theory,
Supergravity (SUGRA), and Twistor Gravity. The details of these
theories need not concern the reader at this point\footnote{The
reader is again referred to the electronic preprints at Los Alamos
for the latest information on Superstring theory and the like:
http://xxx.lanl.gov.}.

The common aspect of all of these theories is that they are
usually associated with some sort of symmetry breaking mechanism,
which in turn gives rise to a phase transition.  In the late
seventies, cosmologists explored the possibility that these
effects may not be negligible \citep{Lindecor1},\citep{Lindecor2}.
In the case of an SU(5) GUT, the model predicted a world dominated
by massive magnetic monopoles. In the early 1980's Guth explored
the possibilities of eliminating these relics and the associated
cosmological consequences, which in turn leads to the concept of
inflation.

One may argue that SU(5) is not known to be the correct theory.
This is true. However, most physicists believe that any correct
unified theory will exhibit symmetry breaking. Moreover, the
electroweak theory has been verified experimentally and exhibits a
symmetry breaking that could have given rise to inflation.

\subsection{Inflation As a Solution to the Initial Value Problems}

It was discussed in the first part of this section that inflation
solves the flatness problem because the universe expands at such a
great rate that the curvature term is `redshifted' away.  Another
way of stating this result is to define inflation as any period in
the evolution of the universe in which the scale factor ($a(t)$)
undergoes a period of acceleration; i.e., $\ddot{a}(t)
> 0$.  This condition can be used to provide a further insight
into what inflation means.  Consider the quantity $(H a)^{-2}$.
Knowing $$H=\dot{a}/a,$$ it follows that $$1/(H
a)^{2}=\dot{a}^{-2}.$$  Now consider the time derivative of this
quantity.  $$\frac{d}{dt} \Big( \dot{a}^{-2} \Big)=-2 \dot{a}^{-3}
\ddot{a}<0,$$ given the conditions $\dot{a}>0$ and $\ddot{a}>0$.
This implies,
\begin{equation}
\frac{d}{dt}\Big( \frac{1}{H^{2}a^{2}} \Big)<0. \label{fix}
\end{equation}
Referring back to equation (\ref{friedconstant}), and dividing
through by $H^{2}$, one again gets the equation for the evolution
of the density parameter, $\Omega=\rho/\rho_{c}$,
\begin{equation}\label{omegainflation}
\mid \Omega(t)-1 \mid=\frac{\mid k \mid}{a^{2}H(t)^{2}}.
\end{equation}
Comparing Equations (\ref{fix}) and (\ref{omegainflation})
expresses the fact that the curvature decreases during inflation.
More explicitly, as $a$ and $H$ increase by tremendous amounts
during inflation, the right had side of (\ref{omegainflation})
approaches zero since the denominator becomes large.  Thus,
$\Omega$ is driven towards one and the universe is made flat by
inflation. As the scale factor evolves under the condition
$\ddot{a}(t)>0$ the density ($\rho$) approaches the critical
density ($\rho_{c}$).

But (\ref{fix}) can also be written as, $d(1/Ha)/dt\;<0$ since $H$
and $a$ are both taken as positive quanitites. Recall that $1/H$
gives the particle horizon of a flat universe, so one can use
Equation (\ref{flatuniverse}),

$$H^{-1}=d_{p}=a(t)r \Longrightarrow 1/Ha =r,$$ where $r$ is the
comoving radial coordinate.  Using $d(1/Ha)/dt\;<0$ gives the
relation, $$\frac{d}{dt}\; (r) < 0.$$

\begin{figure}[]
\includegraphics[totalheight=4.0 in,keepaspectratio]{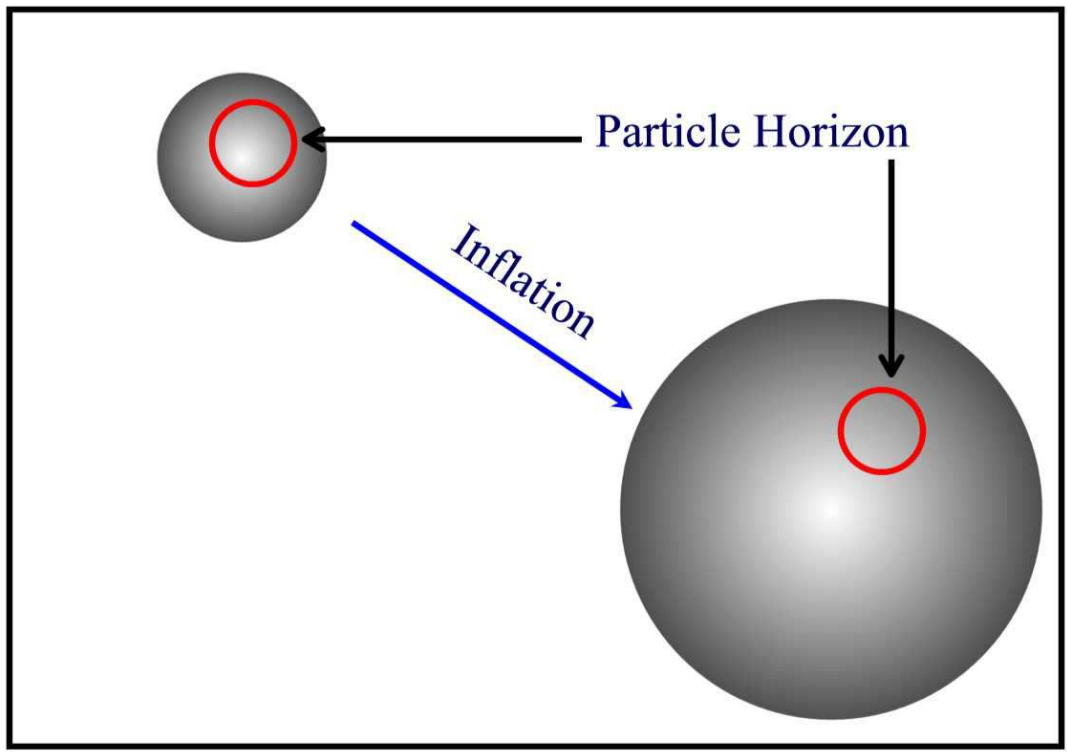}
\caption{This figure illustrates the expansion from the point of
view of an observer located in an embedded dimension.  In this
way, the observer can `look down' and see the expansion take
place.  The spacetime is seen to expand from under the particle
horizon.} \label{figHubble}
\end{figure}

\begin{figure}[]
\includegraphics[totalheight=4.0 in,keepaspectratio]{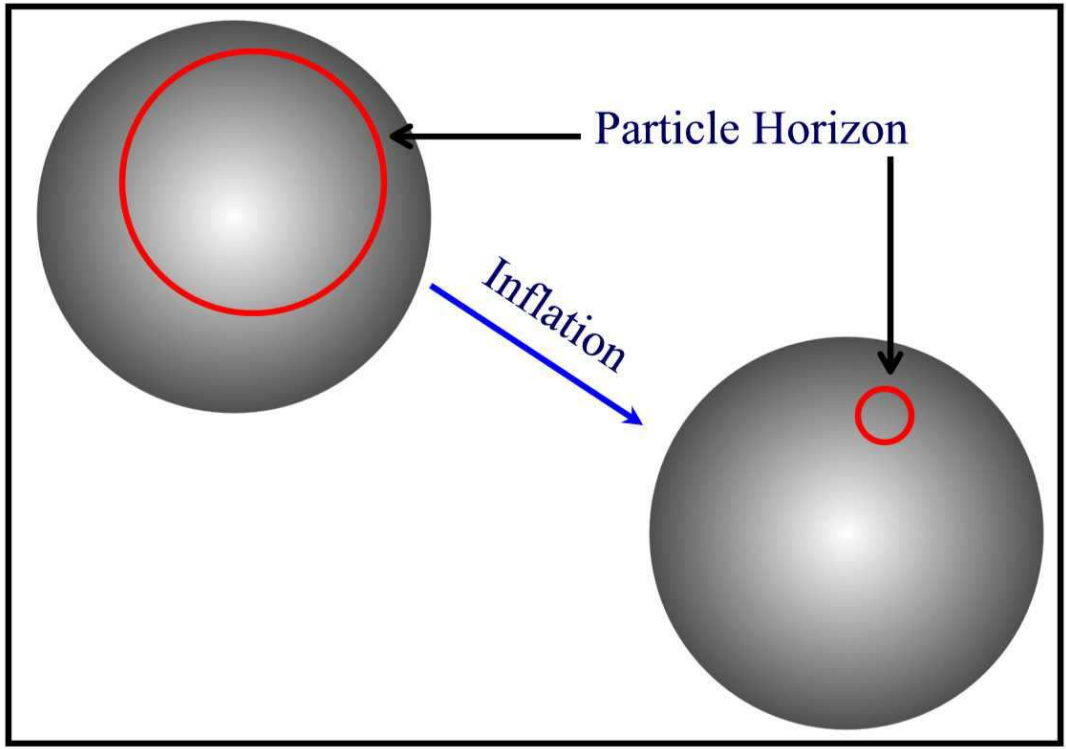}
\caption{This figure illustrates the more natural perspective
contrasted to that of Figure (\ref{figHubble}).  The frame of
reference is that of comoving coordinates, which puts the observer
at rest relative to the expansion.  In this frame the particle
horizon is seen to shrink.} \label{figHubble2}
\end{figure}

What does this mean?  This implies that during a period of
inflation the comoving frame (parameterized by $r,\theta,
\mbox{and} \; \phi$), SHRINKS! Remember that the comoving
coordinates represent the system of coordinates that are at rest
with respect to the expansion.  In other words, instead of viewing
the spacetime as expanding it is equally valid to view the
particle horizon as shrinking.  To visualize this, it is perhaps
useful to again consider the idea of an expanding balloon (see
Figures \ref{figHubble} and \ref{figHubble2}). Normally, in this
example, one views two points on the surface of the balloon as
getting farther apart because the balloon is expanding. However,
if one chooses a frame in which the surface is not expanding this
would mean that the metric, or way of measuring, would shrink.
Thus, the distance between the points would get larger, since the
comoving coordinates got smaller.  Each frame of reference has its
advantages.  For the remainder of this paper I will choose the
frame where the universe is seen to expand.  This has the
advantage that the Hubble length remains `almost' constant during
inflation, which eases the discussion in the analysis to follow.

Notice it is now justified to use the flat universe approximation,
since inflation forces $\Omega=1$ by the fact that $1/a^{2}H^{2}$
increases so rapidly compared to $k$ in Equation
(\ref{omegainflation}). Also note that $\Omega$ doesn't have to be
entirely matter dominated. For example,
$\Omega=\Omega_{M}+\Omega_{\Lambda}=.3+.7=1$ is an acceptable
configuration in the inflation scenario.

\begin{figure}[]
\includegraphics[totalheight=4.0 in,keepaspectratio]{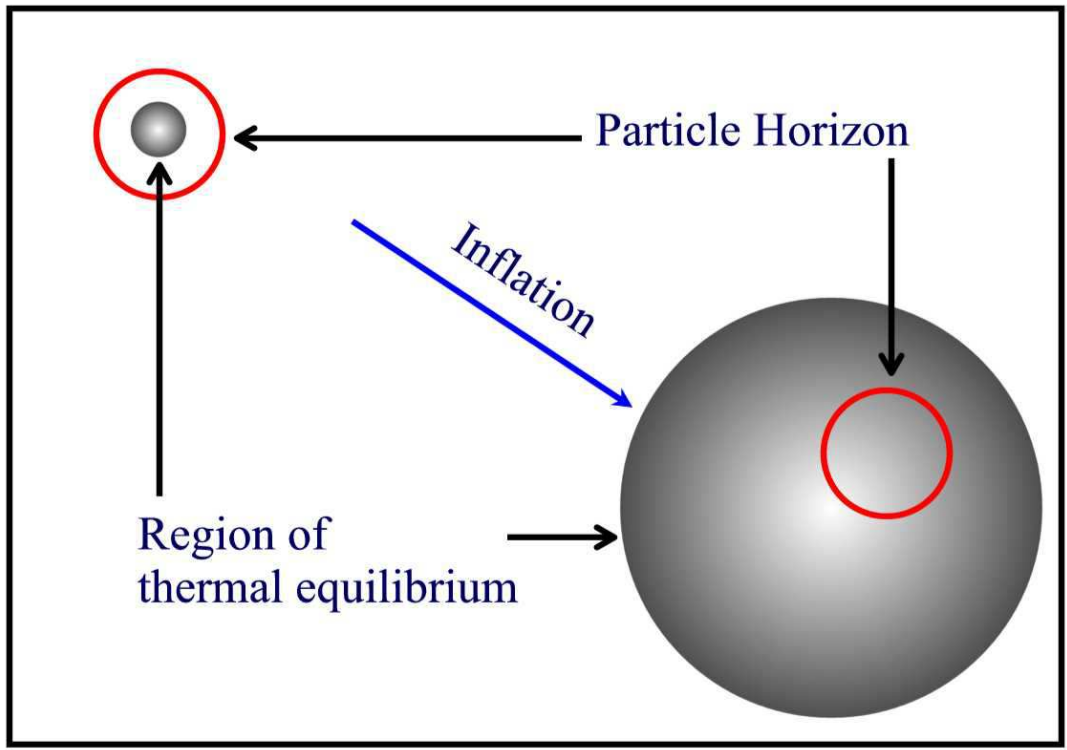}
\caption{This figure illustrates that during inflation regions
that have achieved thermal equilibrium can be expanded outside the
particle horizon (Hubble distance). After inflation, the particle
horizon begins to expand faster than the spacetime and these
regions reenter the horizon.  Inflation is the only known way to
explain this uniformity, thus solving the horizon problem.}
\label{evol}
\end{figure}

So, the picture during inflation is that the spacetime background
expands at an accelerating pace. This resolves the horizon
problem, since causal regions in the early universe are stretched
to regions much larger than the Hubble distance.  This is because
during inflation the scale factor evolves at super-luminal speeds,
whereas the particle horizon (Hubble distance) is approximately
constant.  The particle horizon does expand at the speed of light
(by definition), but this pales in comparison to the evolution of
the scale factor.  Remember the Hubble distance is the farthest
distance light could have traveled from a source to reach an
observer.  Once inflation ends, the scale factor returns to its
sub-luminal evolution leaving the particle horizon to ``catch
up''.  This situation is illustrated in Figure \ref{evol}. So as
we look out at the sky today we are still seeing the regions of
uniformity that were stretched outside the particle horizon during
inflation.

A more quantitative argument is given by considering the physical
distance light can travel during inflation compared to after.

\begin{equation}
\label{stuff1} a(t_{rec}) \int_{t_{inf}}^{t_{rec}} \frac{dt}{a(t)}
\; \gg a(t_{0}) \int_{t_{rec}}^{t_{0}} \frac{dt}{a(t)},
\end{equation} where $t_{inf}$ marks the beginning of inflation,
$t_{rec}$ is the time of recombination, and $t_{0}$ is today.

Equation (\ref{stuff1}) can be understood by making the following
estimates.  In the first integral, the scale factor during
inflation is given by, $a(t) \sim e^{Ht}$.  Whereas, in the second
integral one can assume the scale factor is primarily matter
dominated $a(t)\sim t^{2/3}.$  Furthermore, the integral on the
right can be simplified by taking $t_{rec}=0$. Of course this only
increases the integral.  Lastly, $t_{inf}$ can be set equal to
zero and then $t_{rec}=\Delta t$ is the time inflation lasts.
Thus, $$ e^{H \Delta t} \int_{0}^{\Delta t} \frac{dt}{e^{Ht}} \;
\gg t_{0}^{2/3} \int_{0}^{t_{0}} \frac{dt}{t^{2/3}}.$$ Evaluating
the integrals and a bit of algebra gives,
\begin{equation}
\label{ya1}H^{-1} \Big( e^{H \Delta t} -1 \Big) \; \gg 3t_{0} =
2H^{-1},
\end{equation}
where the last step uses $H=\dot{a}/a=2/3t$.  So, we can see from
(\ref{ya1}) that as long as inflation lasts long enough ($\Delta
t$) then the horizon problem is solved.

With the discussion presented thus far, the monopole problem is
solved trivially.  The number of predicted monopoles per particle
horizon at the onset of inflation is on the order of one
\citep{300years}. As discussed previously, this would result in a
density today that would force $\Omega \gg 1$, which is not
observed.  As stated previously, the comoving (causal) horizon
shrinks during inflation. Thus, if the universe starts with one
monopole, it may contain that one monopole after inflation, but no
more. However, this is highly unlikely if the universe inflates by
an appreciative amount. Furthermore, inflation redshifts all
energy densities.  So, as long as the temperature does not go near
the critical temperature after inflation, no additional monopoles
may form.

This holds true for the other topological defects and unwanted
relics associated with spontaneous symmetry breaking (SSB) in
unified theories. This leads one to ask, why would the temperature
increase after inflation?  The mechanism by which this reheating
of the universe takes place is related to the mechanisms that
bring about the demise of the inflationary period. These
mechanisms are understood through the dynamics of scalar fields,
to be discussed in the next section.

One question has been left unresolved with reference to the
problems of initial values in the hot Big Bang model.  This is the
problem of the origin of structure in the universe.  It was
pointed out that the DeSitter universe is left empty and cold with
no stars or galaxies.  The flatness and monopole problem were
resolved by a redshifting of the various energy densities. But, if
no energy is present, how can particle creation take place? This
peculiar feature of inflation will be discussed in the next
section, but here I would like to present a qualitative
description.

At the end of inflation, all energy densities have become
negligible except the vacuum density (or cosmological constant if
you prefer).  Where did the energy go?  It went into the
gravitational `potential' of the universe, so energy is still
conserved \citep{guthp1},\citep{guthp2},\footnote{Actually, energy
need not be conserved if we live in an open universe.  However,
this need not concern us here.}.  Thus, at the end of inflation
there is a universe filled with vacuum energy, which takes the
form of a scalar field. This scalar field is coupled to gauge
fields, such as the photon. As the scalar field releases its
energy to the coupled field, the universe goes through a reheating
phase where particles are created as in the hot Big Bang model.
The energy for this particle creation is provided by the `latent'
heat locked in the scalar field. More will be said on this later,
but the important point is that the hot Big Bang model picks up
where inflation leaves off. Thus, one may be inclined to say,
inflation is a slight modification to the hot Big Bang model. One
author refers to inflation as, ``a bolt on accessory''
\citep{liddleeprint}.

This all sounds very appealing, however reheating is a fragile
topic for inflation and results in a number of different models.
This derives from the fact that if the temperature is too high at
the time of reheating, the unwanted particle relics could be
re-introduced into the model!  As a result, many different
reheating scenarios have been proposed, along with many different
models for the onset of inflation.

One surprise from inflation makes all of this worry worth it.
Along with offering a solution to the various initial value
problems of the hot Big Bang, inflation offers a mechanism to seed
the large-scale structure of the universe.  Depending on the model
chosen, (e.g., reheating temperature, onset conditions, etc.) one
gets predictions for the large-scale structure of the universe and
the anisotropies in the cosmic background. As will be seen in the
next section, this again demonstrates how the very small (quantum
mechanics) can impact the very large (universal structure). In
some models of inflation, a small fluctuation in the quantum foam
of the Planck epoch ($t<10^{-43}$) can give rise to the formation
of galaxies, solar systems, and eventually human life!  We are the
result of pure chance! This is getting a little ahead of the game,
so let us consider a quantitative and mechanical explanation of
inflation.

\subsection{Inflation and Scalar Fields}
As stated above, inflation is capable of solving many of the
initial value, or `fine-tuning', problems of the hot Big Bang
model. This is assuming that there is some mechanism to bring
about the negative pressure state needed for quasi-exponential
growth of the scale factor. In the early 1980's, Alan Guth
\citep{guth} was studying properties of grand unified theories or
GUTs.  It was found in the late 70's that these theories predict a
large number of topological defects
\citep{Lindecor1},\citep{Lindecor2}. Guth was specifically
addressing the issue of monopole creation in the SU(5) GUT. It was
found that the theory predicts a large number of these monopoles,
and that they should `over-close' the universe
\citep{Lindecor1},\citep{Lindecor2}. This means that the monopole
contribution to $\Omega$ is greater than the observed upper-bound
on the density parameter, $\Omega
>4$, which comes from observation \citep{turner}. To remedy this,
Guth suggested that the symmetry breaking associated with scalar
fields in the particle theory cause the universe to enter a period
of rapid expansion.  This expansion `dilutes' the density of the
monopoles created, as stated above.

The first step in understanding the dynamics of scalar fields is
to undertake the study of field theory.  In field theory, one
considers a Lagrangian density, as opposed to the usual Lagrangian
from classical mechanics. This is because the scalar field is
taken to be a continuous field, whereas the Lagrangian in
mechanics is usually based on discrete particle systems.  The
Lagrangian ($L$) is related to the Lagrangian density
($\mathcal{L}$) by,
\begin{equation}
L=\int \mathcal{L} \; d^{3}x.
\end{equation}
Usually the scalar field is represented by a continuous function,
$\phi(x,t)$, which can be real or complex.  Given a potential
density of the field, $V(\phi)$,  $\mathcal{L}$ takes the form,
\begin{equation}
\label{lagrangedens}
\mathcal{L}=\frac{1}{2}\partial_{\mu}\phi\partial^{\mu}{\phi}-V(\phi).
\end{equation}
The resulting Euler-Lagrange equations result from varying the
action with respect to spacetime \citep{ryder},
\begin{equation}
\frac{\partial{\mathcal{L}}}{\partial{\phi}}-\frac{d}{dx^{\mu}}
\Bigg(
\frac{\partial{\mathcal{L}}}{\partial{(\partial_{\mu}{\phi})}}
\Bigg)=0,
\end{equation}
where $x^{\mu}=(t,-x^{i})$ as usual, and $\hbar=c=1$.  Also note,
$g_{\mu\nu}=\mbox{diag}(1,-1,-1,-1)$, the factor of $\sqrt{-g}$
that usually appears in the action and other equations involving
tensor densities will be $\sqrt{-(-1)}=1$ (Minkowski space). The
resulting equation is
\begin{equation}
\label{scalarmotion} \ddot{\phi}+3H\dot{\phi}=-V^{\prime}(\phi).
\end{equation}
The prime represents differentiation with respect to $\phi$ and
the term containing the Hubble constant serves as a kind of
friction term resulting from the expansion. The field is taken to
be homogeneous, which eliminates any gradient contributions.  This
homogeneity is a safe assumption, since {\em physical} gradients
are related to {\em comoving} gradients by the scale factor,
\begin{equation}
\label{rednabla} \nabla_{{physical}}=a^{-1}(t)\nabla_{{comoving}}.
\end{equation} Thus, the inhomogeneities in the field are
redshifted away during inflation since the scale factor increases
by a large amount.

One can also define the stress-energy tensor by use of Noether's
theorem \citep{ryder},
\begin{equation}
\label{Ttime}
T^{\mu\nu}=\partial^{\mu}\phi\partial^{\nu}\phi-g^{\mu\nu}\mathcal{L}.
\end{equation}
This is useful, because it can be compared to $T^{\mu\nu}$ for a
perfect fluid, namely,

$$T^{\mu\nu}=\mbox{diag}(\rho,p,p,p).$$ Using (\ref{lagrangedens})
in (\ref{Ttime}) yields,

$$\rho=T^{00}=\frac{1}{2}\dot{\phi}^{2}+V(\phi)+\frac{1}{2}(\nabla\phi)^{2}.$$
The calculation for the pressure is a bit more subtle,

$$p=(T^{11}+T^{22}+T^{33})/3.$$

Consider the first component of pressure, again making use of
(\ref{lagrangedens}) and (\ref{Ttime}),

$$T^{11}=\partial^{1}\phi\partial^{1}\phi-g^{11} \Bigg[
\frac{1}{2}\partial_{\beta}\phi\partial^{\beta}\phi-V(\phi)
\Bigg].$$ Since $g^{11}=-1$ and one can use the metric to raise
and lower indices, $$T^{11}=(\partial^{1}\phi)^{2}-V(\phi)+ \Bigg[
\frac{1}{2}g^{\beta
\gamma}\partial_{\beta}\phi\partial_{\gamma}\phi \Bigg].$$ Since
the metric is diagonal this yields,
$$T^{11}=(\partial^{1}\phi)^{2}-V(\phi)+ \frac{1}{2}\Bigg[g^{0
0}\partial_{0}\phi\partial_{0}\phi+ g^{1
1}\partial_{1}\phi\partial_{1}\phi +g^{2
2}\partial_{2}\phi\partial_{2}\phi+g^{3
3}\partial_{3}\phi\partial_{3}\phi \Bigg],$$

$$=(\partial^{1}\phi)^{2}-V(\phi)+
\frac{1}{2}\dot{\phi}^{2}-\frac{1}{2} (\nabla{\phi})^{2}.$$
Similarly, the $T^{22}$ and $T^{33}$ components may be found.  So
for the total pressure one finds, $$p=\frac{1}{3}\sum_{i}T^{ii},$$
or,
\begin{equation}
\label{pressure}
p=\frac{1}{2}\dot{\phi}^{2}-V(\phi)-\frac{1}{6}(\nabla\phi)^{2}.
\end{equation}
From the $T^{00}$ component we already found,
\begin{equation}
\label{density}
\rho=\frac{1}{2}\dot{\phi}^{2}+V(\phi)+\frac{1}{2}(\nabla\phi)^{2}.
\end{equation}

Equations (\ref{pressure}),(\ref{density}) for the pressure and
the energy density, show that the equation, $p=-\rho$ is not quite
satisfied. A first resolution to this problem is to again assume
that the scalar field ($\phi$) is spatially homogeneous, allowing
one to eliminate the gradient terms ($\nabla\phi$). This
assumption is only made at this point to simplify the analysis. As
we have seen if one keeps the terms, it can be shown that the
gradients are redshifted away by the expansion (\ref{rednabla}).
Ignoring gradients, the equations become

\begin{equation}
\label{pressure2} p=\frac{1}{2}\dot{\phi}^{2}-V(\phi),
\end{equation}
\begin{equation}
\label{density2} \rho=\frac{1}{2}\dot{\phi}^{2}+V(\phi).
\end{equation}

The first term $\frac{1}{2}\dot{\phi}^{2}$ can be thought of as
the kinetic energy and the second as the potential, or
configuration energy.  It is now possible to explicitly see where
Equation (\ref{scalarmotion}) came from if we assume the field can
be described as a `perfect' fluid. This assumption allows us to
use a continuity equation, \begin{equation}
\dot{\rho}+3H(\rho+p)=0.
\end{equation}
By plugging in the energy density of the field (\ref{density2})
and making use of the Friedmann equation (to get $H$) one obtains
Equation (\ref{scalarmotion}) in perhaps a more enlightening way.

From the pressure and energy density derived above, we see that
the requirement that $p=-\rho$ can be approximately met, if one
requires $\dot{\phi}\ll V(\phi)$. This leads to what is called the
slow-roll approximation (SRA), which provides a natural condition
for inflation to occur\footnote{In much of the literature on
inflation, the slow-roll approximation is presented as a necessary
and sufficient condition for inflation. However, in many new
models of inflation this is not necessary. For a treatment of
these models, see ~\citep{guthp1},\citep{guthp2},\citep[and
references therein]{lindep1}.}. To assure the constraint on
$\dot{\phi}$, one must also require that $\ddot{\phi}$ be
negligible. Given these requirements, we will to define the
slow-roll parameters and introduce the Planck mass\footnote{The
Planck mass is easier to work with opposed to Newton's constant
$G$, since most of the interesting energy scales are on the order
of GeV ($1 \; eV=1.6 \times 10^{-19} \; \mbox{Joules}$). In these
units, the Planck mass is $\approx 10^{19}$ GeV.}
~\citep{liddle3},

\begin{equation}
\epsilon(\phi)=\frac{M^{2}_{p}}{16 \pi} \Bigg(
\frac{V^{\prime}}{V} \Bigg)^{2}, \label{epsilonSRA}
\end{equation}
\begin{equation}
\eta(\phi)=\frac{M^{2}_{p}}{8 \pi}\Bigg(
\frac{V^{\prime\prime}}{V} \Bigg). \label{etaSRA}
\end{equation}

At this point, it is useful to distinguish $\phi$ as the inflaton
field. Inflaton is the name given to $\phi$, since its origin does
not have to originate with a specified particle theory.  Although
the original hope was that $\phi$ would help determine the correct
particle physics models, current model building does not
necessarily require specific particle phenomenology.  This is
actually an advantage for inflation, it retains its power to solve
the initial value problems, yet it could arise from any arbitrary
source (i.e., any arbitrary inflaton).

However, observation of the large-scale structure of the universe
and anisotropies in the cosmic background should be able to
constrain the inflaton parameters to a particular region.  This
can then be used by particle theorists, as a motivation for some
required scalar field. Observational aspects of inflation will be
considered in the next section, but this property of the inflaton
field manifests itself as one of the greatest contributions of
cosmology commensurate with particle theory. Some examples of
potentials that have been proposed for the inflaton are presented
below \citep{veneziano},\citep{liddleeprint}.

\begin{eqnarray}
& V(\phi)=\lambda(\phi^{2}-M^{2})^{2} \quad & \mbox{Higgs
potential}\\ &V(\phi)=\frac{1}{2}m^{2}\phi^{2} & \mbox{Massive
scalar field}\\ &V(\phi)=\lambda\phi^{4} & \mbox{Self-interacting
scalar field}\\ &V(\phi)=2H_{i}^{2} \Big( 3-\frac{1}{s}\Big)
e^{-\phi / \sqrt{s}} \quad \quad & \mbox{Dilaton scalar field
(string theory)}
%\end{tabular}
\end{eqnarray}

In Guth's original inflation scenario \citep{guth}, the inflaton
field ($\phi$) sits at a local minimum and is trapped in a false
vacuum state (see Figure \ref{fig2}).  The vacuum state of a field
or particle is the lowest energy state available to the system.
Some examples are the ground state of the hydrogen atom (-13.6 eV)
and the ground state of the harmonic oscillator ($\frac{1}{2}\hbar
\omega$). The concept of `false' vacuum comes from examination of
Figure \ref{fig2}.  If $\phi$ is `placed' in the potential well on
the left, the lowest energy available is that of the false vacuum.
The only way $\phi$ can get out of this local minimum is by
quantum tunneling, after some characteristic time. As tunneling
takes place the universe inflates. Inflation halts when $\phi$
reaches the false vacuum and bubbles of the false vacuum coalesce
releasing the `latent' heat that was stored in the field. This is
much like the way bubble nucleation occurs when opening a bottle
of compressed liquid (like soda).  Energy escapes from the soda in
the form of carbon dioxide and the liquid enters a lower more
favorable energy state.

Tunneling that leads to bubble nucleation is a first order phase
transition. This is very similar to processes that take place in
the study of condensed matter physics, fluid dynamics, and
ferromagnetism (see for example \citep{maris} and \citep{kittel}).
The bubbles experience a state of negative pressure.  Once
created, they continue to expand at an exponential rate.  Each
expanding bubble corresponds to an expanding domain. However, when
one carefully investigates this situation, one finds that the
bubbles can collide as they reach the false vacuum.  Furthermore,
the size of these bubbles expands at too great of a rate and the
corresponding universe is left void of structure. One finds that
too much inflation occurs and the visible universe is left empty.
This is referred to as the `graceful exit' problem. Again one is
presented with an empty universe, which was the same reason that
the DeSitter universe idea was abandoned.

Guth and others further tried to remedy these problems by
fine-tuning the bubble formation.  The problem with this is two
fold.  One, cosmologists and particle theorists don't like
fine-tuning.  The idea is to form a model that gives our universe
as a usual result that follows from natural consequences. By
natural one means that the scales of the model are related to the
fundamental constants of nature; e.g., quantum gravity should
occur at the Planck scale, since this scale is the only one
natural in units (c,$\hbar$,G).  Secondly, if the model is
fine-tuned to agree with the observations of the anisotropies in
the cosmic background, the bubbles would collide far too often.
This results in the appearance of topological defects, like the
monopoles.  However, this was the whole reason inflation was
invoked in the first place.

In 1982, a solution to the graceful exit problem was proposed by
Linde \citep{linde6} and independently by Steinhardt and Albrecht
\citep{steinhardt7}. This {\em New Inflation} model solves the
graceful exit problem by assuming the inflaton field evolves very
slowly from its initial state, while undergoing a phase transition
of second order.  Figure \ref{fig3} illustrates this by again
considering the evolution of $\phi$. If $\phi$ `rolls' down the
potential at a slow rate, one obtains the amount of inflation
needed to solve the initial value problems.  After the universe
cools to a critical temperature, $T_{c}$, $\phi$ can proceed to
its `true' vacuum state energy.  The transition of the potential
is a second order phase transition, so this model does not require
tunneling \citep{300years}. This type of transition is similar to
the transitions that occur in ferromagnetic systems
\citep{kittel}.

The majority of current models rely on another concept coined by
Linde as {\em Chaotic Inflation} \citep{lindecor3}.  This model
differs from Old and New Inflation in that no phase transitions
occur. In this scenario the inflaton is displaced from its true
vacuum state by some arbitrary mechanism, perhaps quantum or
thermal fluctuations. Given this initial state, the inflaton
slowly rolls down the potential returning to the true vacuum (see
Figure \ref{chaosinfl}). This model has the advantage that no
fine-tuning of critical temperature is required.  This model
presents a scenario, which can be fulfilled by a number of
different models.

After the displacement of the inflaton, the universe undergoes
inflation as the inflaton rolls back down the potential.  Once the
inflaton returns to its vacuum (true) state, the universe is
reheated by the inflaton coupling to other matter fields.  After
reheating, the evolution of the universe proceeds in agreement
with the Standard Big Bang model.

Although the inflaton could in principle be displaced by a very
large amount, all the inflationist need be concerned with is the
last moments of the evolution.  This is when the perturbations in
the scalar field are created that eventually lead to large-scale
structure and anisotropies in the cosmic background.  As long as
the inflaton is displaced by a minimal amount (minimal to be
defined in a moment) the initial value problems will be solved.
When considering quantum fluctuations resulting in the
displacement of $\phi$, minimal displacement is easily achieved.

Successful evolution is only possible if $V(\phi)$ is very flat
and has minimal curvature. In terms of (\ref{epsilonSRA}) and
(\ref{etaSRA}) this suggests that inflation will occur as long as
the SRA requirements hold.
\begin{equation}
\label{SRAnumbers} \epsilon \ll 1, \;\;\; \;  \;\;\; |\eta| \ll 1.
\end{equation}
This method is successfully used in a number of inflationary
models that make predictions in accordance with observation. It
must be stated again that Chaotic Inflation results in a very
general theory. The inflaton field originally proposed by Guth's
model was that of a grand unified theory, but within the Chaotic
Inflationary scenario any inflaton field can be used that
satisfies the SRA. With these general requirements, potentials
used in supergravity, superstrings, and supersymmetry theories can
be used to motivate inflation.

Using the energy density obtained in equation (\ref{density2}) one
can restate the Friedmann equation (\ref{friedconstant}) in terms
of the scalar field.
\begin{equation}
H^{2}=\frac{8\pi G}{3} \Bigg[
V(\phi)+\frac{1}{2}\dot{\phi}^{2}\Bigg],
\end{equation} Also, the equations of motion derived in (\ref{scalarmotion}) are
restated here for convenience.
\begin{equation}
\ddot{\phi}+3H\dot{\phi}=-V^{\prime}(\phi).
\end{equation} Using the SRA, one can simplify these equations by
eliminating the $\dot{\phi}^{2}$ and $\ddot{\phi}$ terms.  This
leaves the more tractable equations shown below, which remain
valid until $\phi$ approaches the true vacuum; i.e., $\epsilon
\sim 1$.

\begin{equation}
H^{2}=\frac{8\pi V(\phi)}{3 M_{p}^{2}}, \label{SRA1}
\end{equation}
\begin{equation}
3H\dot{\phi}=-V^{\prime}(\phi), \label{SRA2}
\end{equation}
where $M_{p}$ is the Planck mass and has been substituted for $G$,
$$M_{p}\equiv 1/\sqrt{G},$$ remembering that $\hbar=c=1$.

One can use equations (\ref{SRA1})  and (\ref{SRA2}) to manifest
the connection between the slow-roll condition (\ref{SRAnumbers})
and the generic definition of inflation, that is $\ddot{a}>0$.
First note that $$ H=\frac{\dot{a}}{{a}} \Rightarrow \;
\dot{H}=\frac{\ddot{a}}{a}- \Big( \frac{\dot{a}}{a} \Big)^{2}.$$
For inflation to take place means $\ddot{a}(t)>0$ and $a(t)$ is
always positive thus, $$ \frac{\ddot{a}}{a}>0 \Rightarrow \;
\dot{H}+H^{2}
>0,$$ \begin{equation}\Leftrightarrow -
\frac{\dot{H}}{H^{2}}<1. \label{brief} \end{equation} Using
(\ref{SRA1}) and differentiating with respect to time,
$$2H\dot{H}=\frac{8 \pi}{3
M^{2}_{p}}\frac{d\phi}{dt}\frac{d}{d\phi}\Big( V(\phi)\Big),$$
$$\Rightarrow \dot{H}=\frac{8 \pi \dot{\phi} V^{\prime}}{6
M^{2}_{p} H}.$$ Plugging this result into (\ref{brief}) gives,
$$-\frac{\dot{H}}{H^{2}}=-\frac{\dot{\phi}V^{\prime}}{2H V}<1.$$
Solving (\ref{SRA2}) for $\dot{\phi}$ and plugging the result into
the last equation, we obtain
$$\dot{\phi}=-\frac{V^{\prime}}{3H},$$
$$-\frac{\dot{H}}{H^{2}}=\frac{(V^{\prime})^{2}}{6H^{2}V}<1.$$
Lastly, substituting $H^{2}$ from (\ref{SRA1}) one obtains,
\begin{equation}
-\frac{\dot{H}}{H^{2}}=\frac{M^{2}_{p}}{16
\pi}\Big(\frac{V^{\prime}}{V}\Big)^{2}<1.
\end{equation}
But this is just the slow-roll condition $\epsilon \ll 1$. So
again one is reminded that inflation will take place until
$\epsilon \sim 1$, which has been shown to be equivalent to
$\ddot{a}>0$.

\subsubsection{Modeling the Inflaton Field}
The equations of motion derived above (\ref{SRA1}),(\ref{SRA2})
describe the evolution of an arbitrary potential $V(\phi)$ subject
only to the constraint that $V(\phi)$ conform to the slow roll
conditions away from its minimum. As mentioned previously, the
conditions for inflation are arbitrary and inflation will occur as
long as $\ddot{a}>0$. There are three cases that are of particular
interest \citep{peacock}.

\begin{itemize}
\item {\bf Polynomial Inflation.} $V(\phi) \propto \phi^{n}$. This
gives a scale factor which behaves quasi-exponentially. $$a(t)
\sim \exp \Bigg(  \phi^{n/2} kt \Bigg) \;\;\;\; \mbox{where} \;\;
k=\sqrt{\frac{8 \pi}{3 M_{p}^{2}}} .$$

For particle theorist $n=2,4$ are most favorable, since they
describe renormalizable particle theories \citep{ryder}.
\item {\bf Power-law Inflation.}  $V(\phi) \sim \exp \Big( -\sqrt{\frac{16 \pi}{n}} \;
\frac{\phi}{M_{p}}  \Big)$. This gives $a(t) \sim t^{n}$. The only
requirement being that $n>1$.
\item {\bf Intermediate Inflation.}  $V(\phi) \sim \phi^{\mathcal{F}}$, where
$\mathcal{F}=4(n^{-1}-1)$.  This yields $a(t) \sim
e^{(t/t_{0})^{n}}.$
\end{itemize}

As an example consider a simple polynomial model with $n=2$.
$$V(\phi)=\frac{1}{2}m^{2}\phi^{2},$$
$$V^{\prime}(\phi)=m^{2}\phi,$$ $$V^{\prime \prime}(\phi)=m^{2}.$$
The slow roll condition $\epsilon<1$ implies
$$\epsilon=\frac{M^{2}_{p}}{16 \pi}\Bigg( \frac{V^{\prime}}{V}
\Bigg)^{2}=\frac{M^{2}_{p}}{16 \pi}\Bigg(
\frac{m^{2}\phi}{\frac{1}{2}m^{2}\phi^{2}} \Bigg)^{2}<1,$$

$$\Longrightarrow \phi>M_{p}.$$

In other words, the inflaton field must be larger than the Planck
energy.  This is actually the value one would expect for the
cosmological constant, since the only natural scale at high energy
is the Planck scale.  However, inflation (as it has been
presented) relies on the evolution of a classical field.  If the
value of the field is in the quantum regime, where the Planck
scale lives, inflation can not be treated classically.
Fortunately, there are many resolutions to this problem.  First,
what really matters in the field equations is the potential energy
density of the field; namely,
$$V(\phi)=\frac{1}{2}m^{2}\phi^{2}.$$  Remember that $V$ appeared
in the Lagrangian density, thus $V$ is actually a density.  From
this equation one can see that the magnitude of the potential,
and, therefore, the energy scale of the theory, depend on `$m$' as
well as $\phi$.  `$m$' represents the mass of the inflaton and one
resolution to the high value of $\phi$ is to introduce a small
mass for the inflaton; i.e., make `$m$' small.

Another resolution to this scale problem is to limit
considerations to the final part of the evolution of the inflaton.
As stated before, the inflaton evolves until $\epsilon=1$ when the
inflaton then reheats the universe.  The most important
consequences of inflation are its resolution of the initial value
problems and its predictions about large-scale structure and the
cosmic background anisotropies.  As we will see most of these
phenomena only require analysis of the last moments of
`e-foldings' of inflation.  E-folding is a way of measuring how
the scale factor increases. Since, $$a=a_{0} \exp (H \Delta t),$$
one e-folding is defined as the amount of time for $a$ to grow by
a factor of $e$: $$\Delta t=H^{-1}.$$ It will be shown that only
60 e-foldings are needed to resolve the initial value problems and
the scales that are important in determining structure in the
universe only depend on modes that are present during these last
60 e-foldings.

However, it must be pointed out that many people find problems
with these conditions on $\phi$.  The idea of fine-tuning the mass
of the inflaton `$m$' is certainly unappealing. One of the
appealing aspects of inflation was its resolution of the initial
value or fine-tuning problems of the Big Bang model.  But, now we
are again confronted with an initial value problem.  This problem
can be resolved by addressing inflation in the context of a
quantum gravity theory such as superstring theory. I will not
pursue such issues in this paper, although the reader is again
referred to the eprint archive for recent
efforts\footnote{http://xxx.lanl.gov}.

\subsubsection{The Amount of Inflation}
One can find the amount of inflation by considering the change of
the scale factor.  Considering the example of quasi-exponential
expansion, meaning that the Hubble constant need not be constant.
Then,
\begin{eqnarray}
\label{eqnary1} &\displaystyle{a(t)=a_{0} \exp{(H t)}}\nonumber\\&
\displaystyle{\Rightarrow \ln\Bigg(\frac{a(t)}{a_{0}} \Bigg)=H
t},\nonumber\\ &\mathcal{N} \displaystyle{\equiv \ln \Bigg(
\frac{a(t_{final})}{a(t_{initial})} \Bigg)=\int_{t_{i}}^{t_{f}}H
dt \quad \quad }& \mbox{Number of e-foldings}
\end{eqnarray}

Using the slow-roll equations, the number of e-foldings can be
expressed in terms of the inflaton potential. Dividing
(\ref{SRA2}) by (\ref{SRA1}) yields,
\begin{eqnarray}
\frac{3\dot{\phi}}{H}=\frac{3 H
\dot{\phi}}{H^{2}}=-\frac{V^{\prime}}{8 \pi V }\Big( 3 M^{2}_{p}
\Big),\nonumber\\ {\Rightarrow
\frac{\dot{\phi}}{H}=-\frac{M^{2}_{p}}{8\pi} \Bigg(
\frac{V^{\prime}}{V}\Bigg)} \nonumber\\
\end{eqnarray}
Using this result, with the formula for $\mathcal{N}$
(\ref{eqnary1}), one gets,
\begin{displaymath}
\mathcal{N}=\int_{t_{i}}^{t_{f}}H dt=\int_{t_{i}}^{t_{f}}H
\frac{dt}{d\phi} \: d\phi=\int_{t_{i}}^{t_{f}}\frac{H}{\dot{\phi}}
\: d\phi=-\frac{8
\pi}{M^{2}_{p}}\int_{\phi_{i}}^{\phi_{f}}\frac{V}{V^{\prime}} \:
d\phi.
\end{displaymath}

\begin{equation}
\mathcal{N}\simeq \frac{8
\pi}{M^{2}_{p}}\int_{\phi_{f}}^{\phi_{i}}\frac{V}{V^{\prime}} \:
d\phi. \label{efoldSRA}
\end{equation} Here the fact that the SRA has been used is expressed
using `$\simeq$' in (\ref{efoldSRA}).

For $\mathcal{N}>60$, which is needed to solve the initial value
problems \citep{peacock}, we again find $\phi \gg M_{p}$.  This
can be seen from (\ref{efoldSRA}), where $V^{\prime} \sim V/\phi$
using the SRA. This means that if one chooses a potential of
$\frac{1}{2}m^{2}\phi^{2}$, one must choose the coupling, $m^{2}$
to be small.  Given a self interacting potential term, $\lambda
\phi^{4}$, the coupling must be extremely weak, $\lambda \ll 1$.
This coupling agrees nicely with theories of supergravity and
certain string theories, although other potentials are ruled out
because of their couplings, such as the weak coupling.  This
leaves the question. Can inflation be considered without a theory
of quantum gravity? As mentioned before, the inflationist is often
not concerned with these initial stages of inflation.  The common
standpoint is that chaotic inflation can present an evolution and
then one studies the predictions of this evolution. As
aesthetically displeasing as this may be, it allows cosmology to
progress further without a quantum theory of gravity. Ultimately
this issue will be addressed within the context of a theory of
quantum gravity to create a complete picture of the creation of
the universe.  However, it has been argued that a complete
understanding of the universe may be avoided in a scenario known
as eternal inflation \citep{Lindecor4},\citep[and references
within ]{guthp2},\citep[and references within]{guthp1}.  Time does
not permit to discuss these models in detail, however for a
popular account \citep{Lindecor5} is an excellent starting point.

Given the slow-roll conditions and the number of required
e-foldings ($\mathcal{N}$), one can test a model inflaton to see
if it is compatible with an inflationary scenario. With this
generic framework that has been set forth, one can construct
particle theories and then test their validity within the context
of inflation theory.  However, the slow-roll approximation and
initial value problems ($\mathcal{N}>60$) are not the only
constraints on the inflaton and therefore particle theory.  The
inflaton is further restricted by the predicted large-scale
structure of the universe, along with the mechanisms involved with
reheating of the universe at the end of inflation. The large-scale
structure is determined by density perturbations resulting from
quantum fluctuations in the evolution of the inflaton field.  This
analysis can actually be done without the advent of quantum
gravity; however, it is outside the scope of this paper.  Instead,
the author hopes to manifest the stringency of these parameters on
the inflaton field by addressing the observational consequences
and predictions that inflation offers. In the next section these
observational tests will be explored.
%------------------------------------End Section Four-------------------------
\newpage
%------------------------------------Begin Section Five-----------------------
\section{Observational Tests of Inflation}

It was shown in the last section that if the number of e-foldings
exceeds $60$, then inflation can solve the horizon, flatness, and
relic (monopole) problems.  One generally favors this model over a
Big-Bang model, because of its naturalness.  That is to say,
inflation offers a generic scenario for solving the initial value
problems.  However, this arbitrariness can also be viewed as a
problem for inflation.  For instance, throughout this paper it has
been assumed that inflation necessarily leads to a flat universe
($k=0$). However, Hawking, Turok, Linde, and others have shown
that inflation can result in a non-flat universe
\citep{open1},\citep{open2}.  Models can be created that produce
unwanted or wanted relics and contain inhomogeneities.
Furthermore, we have seen inflation requires an inflaton field to
drive the inflation. Where does it come from and what is its
natural value? Originally Guth had the inflaton as corresponding
to a GUT transition; however, today the preferred energy range is
on the order of the Planck scale. Thus, to fully understand
inflation one needs a full quantum theory of gravity. For these
reasons one may ask; Is inflation a particular type of
cosmological model, or is inflation an arbitrary constituent of
any successful theory of the cosmos?

Inflation's strength today can be seen from its predictions of
large-scale structure. Different models predict different
structure and this can be used to narrow the number of possible
models.  One can further constrain the inflationary models by
cosmological parameters. The cosmic background observations,
galaxy surveys, lensing experiments, and standard candles can be
used pin-down the cosmological parameters. In this way,
observational parameters (e.g., $H$,
$\Omega_{M}$,$\Omega_{\Lambda}$) can be given viable ranges and
inflationary parameters can be determined based on these preferred
ranges. With the cosmological parameters determined, inflation
parameters depend only on the height and shape of the inflaton
potential. The inflaton potential correspondingly yields
predictions about the large-scale structure of the universe and
the anisotropies in the cosmic background radiation.

The study of large-scale structure has been pursued for many years
\citep{zeldovich}.  The problem was that there was an appealing
mechanism which could produce the types of perturbations needed to
produce the observed large-scale structure.  These perturbation
types are manifested by the anisotropies in the cosmic background.
The anisotropies result from acoustic oscillations in the
baryon-photon fluid just before recombination. Therefore, the
anisotropy spectrum offers a `snap-shot' of the seeded
inhomogeneities that eventually resulted in galactic structure.
These inhomogeneities were first discovered when COBE mapped the
cosmic background in the early 1990's \citep{smoot}. In this
intimate way, the cosmic background and galaxy surveys predict a
scheme by which structure was formed. The type of perturbation
that is needed only results from models which predict Gaussian,
adiabatic, nearly scale-invariant perturbations \citep{carroll42}.
The only known models that fall into this category are the
inflationary models
\citep{guth},\citep{linde6},\citep{steinhardt7}.

\subsection{Perturbations and Large-Scale Structure}

During the inflation epoch, perturbations (small fluctuations) of
two types are generated: scalar (density) perturbations, and
tensor (metric) perturbations. The scalar perturbations come from
quantum fluctuations of the inflaton field before and during its
evolution.  Tensor perturbations arise from quantum fluctuations
in the space-time metric within the quasi-DeSitter spacetime.
Fluctuations of this type are a natural consequence of a DeSitter
spacetime.

In DeSitter space there exists an event horizon. Consider the
distance light can travel in comoving coordinates, which is given
by (\ref{cosmicdist}) and $a(t) \sim \exp(Ht)$.
\begin{equation}
r=\int_{0}^{\infty} \frac{c \; dt}{a(t)}=\int_{0}^{\infty}
\exp(-Ht) c \; dt=c/H.
\end{equation} The presence of this event horizon suggests the
presence of thermal fluctuations in the fields, similar to those
present in a black hole.  This can be understood by appealing to
the uncertainty principle.  The event horizon causes the ground
state modes of any fields present to be restricted in spatial
extent.  The uncertainty principle then requires, $\triangle p
\geq \hbar H/c$, since the characteristic size is just $c/H$. This
uncertainty in momentum gives rise to energy fluctuations and the
corresponding Hawking temperature is given by \citep{peacock},
\begin{equation}
\label{hawking} kT_{DeSitter}=\frac{H}{2\pi},
\end{equation} where $k$ is the Boltzman constant relating the
energy to the temperature.

This result provides a motivation for the existence of
fluctuations in the metric and scalar field. As these
perturbations are created during inflation they are inflated
outside of the causal horizon (particle horizon). As mentioned in
the previous section, the causal horizon is nearly stationary
during inflation. Once the perturbation has been inflated outside
the horizon, its ends are no longer in causal contact. In this
way, the perturbations become `frozen-in' as classical
perturbations.

Ignoring the nonlinear, or super-horizon, effects of these
perturbations may trouble the reader. However, as we shall see,
ignoring these effects appears to be in agreement with the cosmic
background data \citep{hu2}, \citep{hu3}, \citep{hu4}. As stated,
one reason for choosing to ignore super-horizon evolution is the
causal separation of the ends of the perturbation.  However, this
argument is far from rigorous and the study of nonlinear
perturbations takes much care. Perturbation evolution relies on
the extrinsic (super-horizon) properties of the spacetime manifold
and is sensitive to the gauge of general relativity.  The
perturbation evolution can usually be ignored in regions where the
pressure becomes negligible, which happens to be on the order of
the horizon \citep{peacock}.  This generally motivates one to
ignore the super-horizon evolution, however for a complete
treatment, see \citep{branden},\citep{ref31},\citep[Chapter 8 and
9]{turner}.

After inflation the expansion continues at sub-luminal speeds and
these perturbations enter back inside the causal horizon. Thus,
the most important perturbations for creating structure come from
the ones that were exited near the end of the inflationary period
(i.e., approximately the last 60 e-foldings). After the reheating
process occurs, the inhomogeneities passing back inside the
horizon cause fluctuations that seed the large-scale structure.

Pure exponential inflation, which corresponds to a DeSitter
spacetime, has an interesting property. The spacetime is invariant
under time translation. That is to say, there is no natural origin
of time under true exponential expansion \citep[Chapter
11]{peacock}.  The only fundamental size in the theory is that of
the Hubble horizon ($c/H$).  Thus, one expects that the amplitude
of a `standing wave' perturbation will be related to the horizon
size, $c/H$, which is not changing.  Therefore, we see why
inflation predicts a scale-invariant spectrum for the
perturbations.

This analysis can be illustrated through musical analogy. The
fundamental mode of the perturbations are determined by the Hubble
distance ($c/H$), much like the fundamental mode of a flute is
determined by its length. Because the Hubble length (horizon) is
{\em nearly} stationary during inflation, this means inflation
predicts a scale-invariant, or Harrison-Zeldovich spectrum.
Furthering this analogy, the `overtones` of the universe
correspond to the inflaton potential that determines its behavior,
much like overtones can be used to distinguish one instrument from
another. However, as we have seen that inflation need not be
exponential. The small deviations from DeSitter spacetime result
in small deviations from a scale-invariant spectrum.  These
deviations can be used to successfully predict the correct
potential for the inflaton.

The generated perturbations can be characterized by a power
spectrum, $\delta_{H}^{2}(k)$. The $H$ indicates that the
perturbation amplitude is taken to correspond to its value when it
crossed the causal horizon.  Quantitatively, this corresponds to
$k=aH$, where $k$ is the wave number of the perturbation. Quantum
field theory can be used to calculate an expression for
$\delta_{H}$ similar to Equation (\ref{hawking}) above
\citep[Chapter 8]{peacock}.

\begin{equation}
\label{yap2} \delta_{H}=\frac{H^{2}}{2\pi \dot{\phi}},
\end{equation} where $\phi$ is the inflaton.  This formula manifests
the connection between the inflaton potential and the
perturbations generated during inflation.

One is generally interested in the scale dependence of the
spectral index of these perturbations, since this dependence
changes for different inflation models \citep{lyth21}.

\begin{equation}
n(k)-1 \equiv 2 \frac{d \delta_{H}}{d\ln{k}}. \label{kone}
\end{equation}

For an absolute scale-invariant spectrum, it follows that $\delta$
is independent of $k$ and the above relation gives $n=1$ as one
would expect.  For a spectrum that is nearly scale-invariant the
amount $n$ differs from one is referred to as the tilt of the
spectrum.  Although it is not at all obvious, (\ref{yap2}) and
(\ref{kone}) can be used along with the slow roll parameters
(\ref{etaSRA}),(\ref{epsilonSRA}) to express the tilt as,
\begin{equation}
1-n=6\epsilon-2\eta.
\end{equation} The details of the calculation need not concern us here,
for a derivation of this result see \citep{dekel} or
\citep{peacock}.  This relation is only presented to demonstrate
that the tilt, which is a discriminating factor between models,
can be written in terms of the SRA parameters $\epsilon$ and
$\eta$.  Therefore, if an experimental consequence of the tilt is
observable, one can find the appropriate values for the SRA and
reconstruct the inflaton potential \citep{lyth34}.

\subsection{The Cosmic Background Anisotropies}
The discovery of the anisotropies in the cosmic background by COBE
created a new opportunity for verification of cosmological
parameters and theories of large-scale formation.  As discussed at
the beginning of this paper, the CMB offers a `snap-shot' of the
universe at the time of recombination ($z \sim 1000$).  The
anisotropies that were present in the baryon-photon plasma at this
time are manifested today by the temperature fluctuations in the
spectrum.  These fluctuations are representative of a nearly
Gaussian, scale-invariant spectrum.  As discussed previously, this
is a unique prediction of inflation theories.

Although the quantitative details can become formible, the
qualitative description of these temperature fluctuations is quite
simple.  During inflation, the perturbations formed must be of
nearly the same amplitude and randomly distributed, as discussed
above. After reheating takes place, these classical perturbations
re-enter the horizon causing density fluctuations in the
baryon-photon plasma.  In over-dense regions, potential wells form
that trap the plasma and cause it to heat up.  At the same time,
photon pressure induces a kind of restoring force to oppose the
gravitational potential.  In this way, a harmonic oscillator
motion is set up in the plasma.  These oscillations continue with
no friction (viscosity) from the fluid.  This is why they are
referred to as adiabatic fluctuations.  However, if this were not
the case and the friction is deemed important, one obtains an
isocurvature spectrum \citep{huhuhu}.  These turn out to be
indicative of cosmic strings, which are ruled-out by observation
as a method of primordial structure formation.  However, models
containing cosmic strings that are produced during reheating
following inflation may still play a major role in cosmological
models \citep{brandenberger}.

At the time of recombination, when the photons were able to escape
the fluid, they had to overcome the gravitational potentials. The
picture is that the photons in these potential wells were hotter
than the average, but this temperature difference was partially
cancelled by the gravitational redshift resulting from the photons
`climbing' out of the potential well.  This phenomena is know as
the Sachs-Wolf effect.  The result is that the photons that were
in the wells have a slight temperature increase from those that
were not. This variation is predicted by theory to be on the order
of $10^{-5}$ \citep{peacock}.  These oscillations propagate
through the fluid at the speed of sound.  Thus, there is a
acoustic horizon that is generated within the surface of last
scattering and if present today would have an angular size of
about one degree on the sky.

One concern with the simplicity of this analysis is what effects,
such as reionization in the surface of last scattering, must be
considered? It is important to consider the mean free path of the
photon as it travels within the fluid before escaping. This could
affect the energy and therefore temperature of the spectrum.
However, it turns out that this effect only appears on small
angular scales within the spectrum and can be ignored \citep{hu3}.
Also, any effects from CMB scattering off interstellar gas only
appear in the spectrum at very small angles.  These observations
are of course useful, but offer little insight into examination of
the early universe and formation of large-scale structure.

Given the predicted anisotropies from the Sachs-Wolf effect, the
next step is to examine the cosmic background spectrum through
observation. The anisotropies in the temperature of the cosmic
background spectrum can be expanded in spherical harmonics
\citep{carroll42},
\begin{equation}
\frac{\Delta T}{T}=\sum_{lm} a_{lm} Y_{lm}(\theta,\phi),
\end{equation} The multipole coefficients are
given by, \begin{equation} a_{lm}=\int dA \: Y^{*}_{lm}
\frac{\Delta T}{T}.
\end{equation} The amount of anisotropy at multipole moment $l$ is
expressed by the power spectrum,
\begin{equation}
C_{l}=| a_{lm}|^{2}.
\end{equation}
The $C_{l}$'s measure the temperature anisotropy of two regions
separated by angle $\theta$.  This angle is related to the $l$'s
by: $\theta \sim 180^{^{o}}/l.$ Thus, $l$ allows one to express
the temperature variations of regions separated by an angle
$180^{^{o}}/l$.  The $l=0$ represents the monopole contribution to
the anisotropy, which is of course zero (this is comparing a point
separated from itself by $360^{^{o}}$. The next moment, $l=1$, is
the dipole moment, which compares regions separated by
$180^{^{o}}$. This anisotropy originates from our peculiar
velocity, the motion of the Earth relative to the cosmic
background. This moment is usually taken out of the spectrum to
leave the `true' anisotropy.  The $l=2$ moment is the quadrupole
contribution, which marks the first non-trivial anisotropy for
understanding structure formation.

When the COBE data is plotted with multipole $l$ versus
temperature variation, a peak is found to occur around $l=220$ or
$1^{^{o}}$ (Figure (\ref{cmb})). This peak corresponds to the
angular size on the sky of the acoustic horizon discussed before
and has been called the Doppler peak. Thus, there is a maximum
temperature variation at precisely the angle predicted by the
Sachs-Wolf effect. Since a mechanism of this type can only be
explained by an inflationary model, one is presented with a strong
argument for inflation \citep{huhuhu}.

However, the peak is actually sensitive to the cosmological
parameters, such as the Hubble constant and the curvature of the
universe.  If the universe is non-flat then the null geodesics are
found to converge (diverge) in the case of spherical (hyperbolic)
geometry. The angle subtended is given by $\theta_{H} \sim
\Omega^{1/2} \: 1^{^{o}}$. This means for the peak at $l=220$ we
live in a universe which is flat.  Although, this seems to
represent a bit of circular logic.  This difficulty can be
remedied by calling upon other observational tests to constrain
the parameters.  These includ galaxy surveys, lensing experiments,
or standard candle observations.  When all of these methods are
combined, strong constraints can be put on parameters and the best
model can be determined.

The cosmic background is apparently richer in structure than was
first realized.  As we have discussed, inflation predicts
fluctuations in both the scalar and tensor fields.  This gives
rise to slight differences in the anisotropies at small angles
(large $l$). Also, there are multiple peaks in the spectrum
following the peak at $l=220$ and the height and shape of the
spectrum are related to the specifics of inflation models, such as
the tilt.

To examine one aspect of the complexity involved, consider that
the scalar and tensor perturbations can be fixed by their
contribution to the quadrupole moment $C_{l=2}$ of the CMB
\citep{kamionkowski}.
\begin{equation}
\mathcal{S}=6 \: C^{scalar}_{l=2} = 33.2 \:
[V^{3}/(V^{\prime})^{2}],
\end{equation}
\begin{equation}
\mathcal{T}=6 \: C^{tensor}_{l=2} = 9.2 V,
\end{equation} where $V=V(\phi)$ is the inflaton potential, $\mathcal{S}$ is the
scalar contribution, and $\mathcal{T}$ is the tensor contribution.
When the slow-roll approximation is considered and the
determination of cosmological parameters is found by methods other
than CMB analysis; e.g., for large-scale structure probing, one
finds that the ratio $\mathcal{T/S}$ is less than order unity.
This restricts $V\leq 5 \times 10^{-12}$.  Reformulating equation
(\ref{kone}) in terms of the inflaton potential, one finds in
Planckian units the relations for the scalar and tensor spectral
indices, respectively.

\begin{equation}
1-n_{s}=\frac{1}{8\pi}\Bigg(\frac{V^{\prime}}{V} \Bigg)^{2} -
\frac{1}{4 \pi},
\end{equation}
\begin{equation}
n_{t}=-\frac{1}{8 \pi} \Bigg( \frac{V^{\prime}}{V} \Bigg)^{2}.
\end{equation}

Although most models of inflation predict a near scale-invariant,
or Harrison-Zel'dovich spectrum, the small deviations from
differing potentials $V(\phi)$ give a way of testing inflation
models.  With future experiments such as
MAP\footnote{http://map.gsfc.nasa.gov} and
PLANCK\footnote{http://astro.estec.esa.nl/SA-general/Projects/Planck},
these spectral indices will be found with great precision and the
shape and height of the spectrum can be used to manifest the
correct inflaton potential. In this way, inflation will be used to
predict new particle physics, instead of the original scenario
which was vice versa.

However, the ultimate test of inflation is the precise
determination of the tensor perturbations, that is the $n_{t}$'s.
This can't be deduced from the CMB spectrum because both the
scalar and tensor perturbations contribute to the temperature
anisotropy. However, if a method could be devised to separate out
the tensor perturbations and this spectrum were detected, it would
be concrete evidence of an inflationary period. This is because
inflation is the only way metric perturbations can survive to the
present.  This is due to the structure of a DeSitter spacetime.

The tensor spectrum can be separated by creating a polarization
map of the CMB.  The separation is then possible because the
tensor perturbations have an intrinsic axial component, and the
angular dependence can be determined from the metric. Whereas, the
scalar perturbations have no dependence on direction. Therefore,
the polarization vector can be constructed out of two parts, a
curl and a gradient.
\begin{equation}
\vec{P}(\theta,\phi)= \vec{\nabla}A + \vec{\nabla} \times \vec{B},
\end{equation} where $\hat{n}$ gives the direction. Thus, to obtain the tensor terms
one can take the divergence of this vector.  Before proceeding
further, it may be of interest to the reader why the perturbations
only contain tensor and scalar contributions and vector type
perturbations are absent. This is because massless vector fields
are conformally invariant \citep{turnertests}. This invariance can
be broken by introducing a mass term or by explicitly breaking the
coupling. Although for a massive vector field the perturbations
generated are far too small and die off far too quickly during the
inflationary expansion \citep{liddletests}. Thus, a standard
prediction of inflation is that there are no vector perturbations.
But, isn't the polarization a vector? No.  The polarization is
actually a $2 \times 2$ trace free symmetric tensor.  This tensor
is written in terms of the Stokes parameters $Q(\hat{n})$ and
$U(\hat{n})$, which give us the polarization in each direction.
For an explanation of these parameters see, \citep[section
7.2]{jackson}.
\begin{equation}
   \mathcal{P}_{ab}(\hat{n})=\frac{1}{2} \left( \begin{array}{cc}
   Q(\hat{n}) & -U(\hat{n}) \sin\theta \\
   -U(\hat{n})\sin\theta & -Q(\hat{n}) \sin^{2}\theta \\
   \end{array} \right).
\label{whatPis}
\end{equation} The polarization tensor can be expanded in tensor spherical
harmonics \citep{ref14},

\begin{equation}
\frac{\mathcal{P}_{ab}(\hat{n})}{T_{0}} = \sum_{lm} \Bigg[
a^{G}_{(lm)}Y^{G}_{(lm)ab}(\hat{n}) +
a^{C}_{(lm)}Y^{C}_{(lm)ab}(\hat{n}) \Bigg]
\end{equation}

The $Y^{G}_{(lm)ab}$ and $Y^{C}_{(lm)ab}$ represent a basis for
the gradient (scalar) and curl (tensor) perturbation terms in this
polarization mapping.  The $a^{G}_{(lm)}$'s and $a^{C}_{(lm)}$'s
are again just found by exploiting the orthogonality.  One can use
this spectrum to construct necessary requirements for the
potentials of the inflaton field and this analysis can therefore
be used to distinguish the various inflation models.

\subsection{Summary}
Scalar perturbations are the most easily detected form of
perturbation. The scalar nature of these fluctuations arise from
the fact that the perturbations are of mass fields in the
primordial era, that is, before the time of decoupling. Different
models of inflation and the corresponding reheating mechanisms
differ in their predictions of mass variation. Regardless, these
variations of the early universe eventually give rise to the
structural formation of galaxy clusters, which can help
distinguish the various theories of inflation and structure
formation. Tensor perturbations are also detectable and these
perturbations result from fluctuations in the space-time metric in
the primordial era. Again, these perturbations are very small and
would be very hard to detect.  However, certain models of
inflation predict wavelengths that could be detected by laser
interferometry gravitational wave detectors, such as
LIGO\footnote{http://www.ligo.caltech.edu/}. If these waves were
detected it would help eliminate many inflation models and help
narrow the region of viable theories. Furthermore, inflation is
the only theory that can currently account for a gravitational
wave spectrum.  The detection of the spectrum would be a great
success for the inflation theory.

Both of these types of perturbations contribute to the $10^{-5}$
temperature fluctuation in the cosmic background.  The biggest
challenge for experimentalists is to separate the scalar and
tensor contributions to the temperature fluctuations.  In practice
this is very difficult, if not impossible, and it becomes more
practical to consider the polarization of the CBR.

For the inflationist, the goal of CRB measurements is to
distinguish between the various models of inflation.  A good way
to begin, is to express many of the relations obtained thus far,
in terms of $\epsilon$ and $\eta$.  The number of e-foldings
($\mathcal{N}$) can be expressed in terms of $\epsilon$ using
(\ref{epsilonSRA}) and (\ref{efoldSRA}),

\begin{equation}
\label{Nepsilon} \mathcal{N}=\frac{2 \sqrt{\pi}}{M_{p}}
\int^{\phi_{f}}_{\phi_{i}} \frac{d\phi}{\sqrt{\epsilon}}.
\end{equation}

Another useful relation, which may be found in the literature
\citep{ref31}, gives a measure of when a given perturbations of
wave length $k$ passes through the horizon and is therefore
`frozen out'.  This can be expressed as the number of e-foldings
$\mathcal{N}(k)$ from the end of inflation.

\begin{equation}
\mathcal{N}(k)=62-\ln{\frac{k}{a_{0} H_{0}}}-\ln{\frac{10^{16} \:
\text{GeV}}{V^{1/4}_{k}}}+\ln{\frac{V^{1/4}_{k}}{V_{e}^{1/4}}}-\frac{1}{3}\ln{\frac{V_{e}^{1/4}}{\rho^{1/4}_{RH}}}.
\end{equation} $V_{k}$ is the potential when the mode $k$ leaves the horizon,
$V_{e}$ is the potential at the end of inflation, and $\rho_{RH}$
is the energy density after reheating.  This expression may appear
formidable, however it can be used to begin understanding density
fluctuations.  For example, the modes $k$ entering the horizon
today, left the horizon at $\mathcal{N}(k)=50-70$.  The
uncertainty in this range manifests the lack of knowledge of the
inflaton potential.  Thus, once again different inflaton models
make different predictions.

With the rapid advances in observational cosmology, cosmologists
are able to use the abundance of data that is being obtained by
the Hubble Space Telescope, balloon experiments, satellites (such
as Chandra), etc. to narrow the parameters of the universe.  Then
with these values and the relations that have been presented in
this section, one can use inflation to predict new physics for the
pre-inflation or Planckian epoch.  Ultimately this physics will
need a quantum theory of gravity or supersting theory, but
determination of the inflaton potential and the resulting
large-scale structure will set stringent limits in which to test
the predictions of these new theories.  In this way, inflation
offers the link between the innerspace of the quantum realm and
the outerspace of the large-scale structure of the universe.  The
marvelous universe in which we live, the beauty that surrounds us,
and even ourselves, will be the result of a quantum fluctuation or
perhaps a chaotic mishap.
%------------------------------------End Section Five ------------------------
\newpage
%________Begin Bibliography___________________________________________________
\bibliographystyle{apsrev}

%________End Bibliography_____________________________________________________

\appendix
\newpage
%-------------------------------Begin Acronyms--------------------------------
\section{Acronyms}
To hopefully ease the burden of dealing with the ubiquitous
acronyms of cosmology, I have compiled a list of the most common
to be encountered in this paper.

\begin{itemize}
\item CBR --  Cosmic Background Radiation
\item CMBR/CMB -- Cosmic Microwave Background Radiation
\item COBE -- Cosmic Background Explorer
\item CP -- Cosmological Principle
\item CRF -- Cosmic Rest Frame
\item eV,MeV -- electron Volt, Mega-electron Volt
\item MAP -- Microwave Anisotropy Probe
\item QSO's -- Quasi Stellar Remnants
\item RWM -- Robertson Walker Metric
\item SRA -- Slow Roll Approximation
\item SR -- Special Relativity
\item VEV -- Vacuum Expectation Value
\end{itemize}
%-------------------------------End Acronyms ---------------------------------
\newpage
\section{Figures and Tables}
\listoffigures \listoftables
%-----------------------------Begin Figures-----------------------------------

\begin{figure}[]
\includegraphics[totalheight=4.0 in,keepaspectratio]{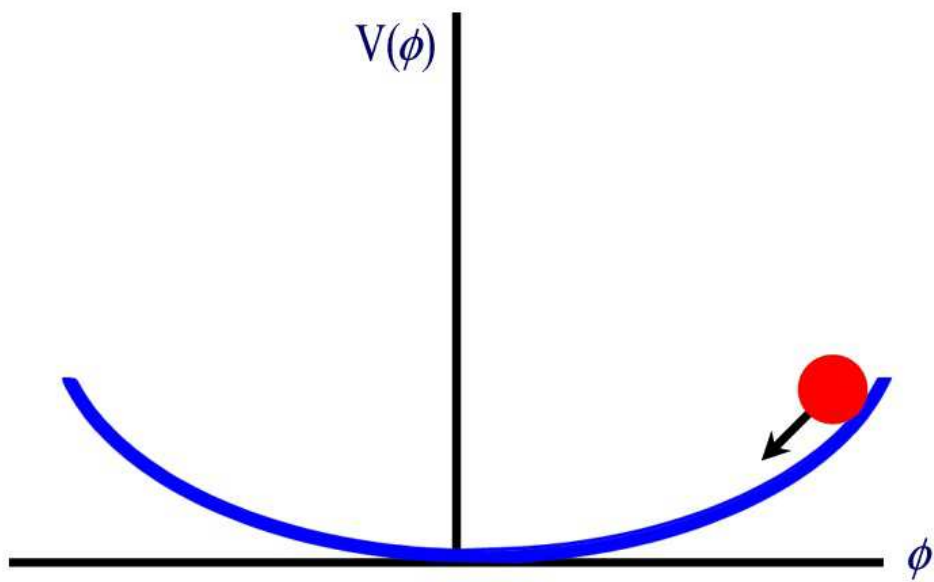}
\caption{Evolution of the Inflaton Field} \label{fig1}
\end{figure}

\begin{figure}[]
\includegraphics[totalheight=4.0 in,keepaspectratio]{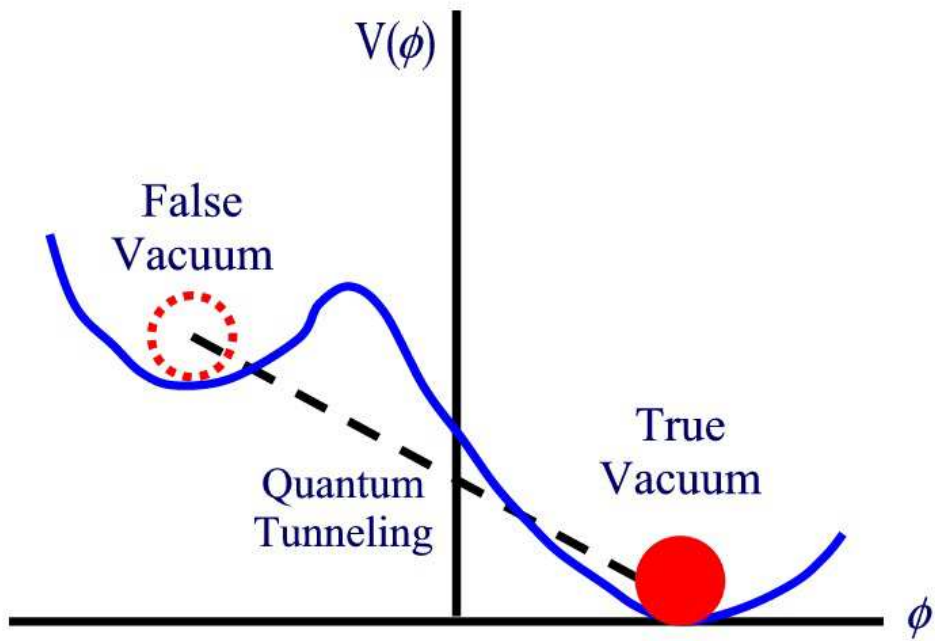}
\caption{Toy Model For Original Inflaton} \label{fig2}In this
model of inflation the inflaton finds itself trapped in a false
minimum.  It is freed from this minimum when tunneling is allowed
to occur resulting in a first order phase transition in the early
universe.
\end{figure}

\begin{figure}[]
\includegraphics[totalheight=4.0 in,keepaspectratio]{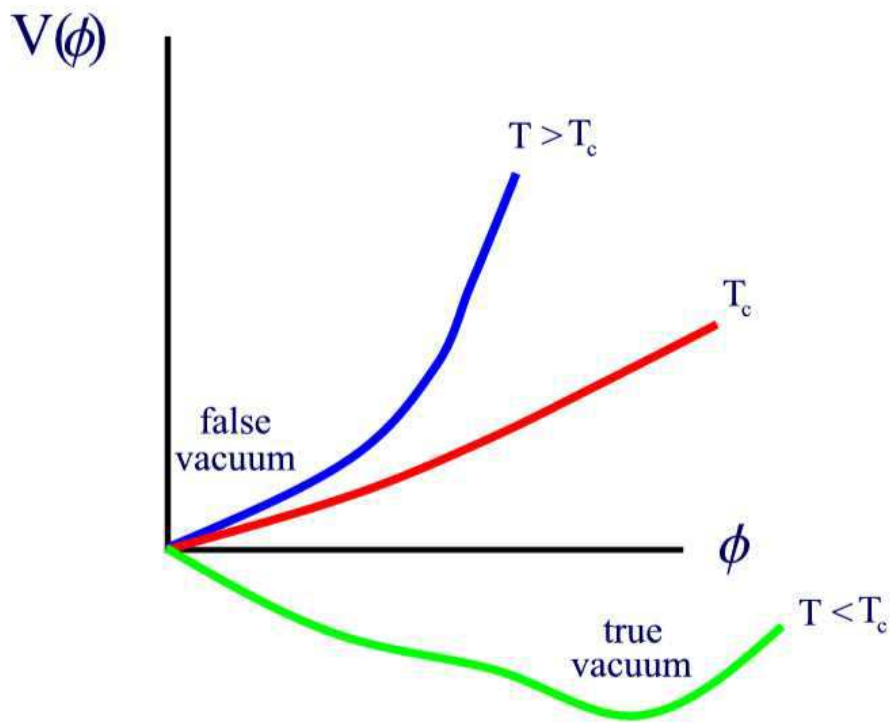}
\caption{Toy Model For New Inflation -- When the temperature of
the universe decreases to the critical temperature $T_{c}$, the
scalar field potential experiences a second order phase
transition.  This makes the `true' vacuum state available to
$\phi$. } \label{fig3}
\end{figure}

\begin{figure}[]
\includegraphics[totalheight=4.0 in,keepaspectratio]{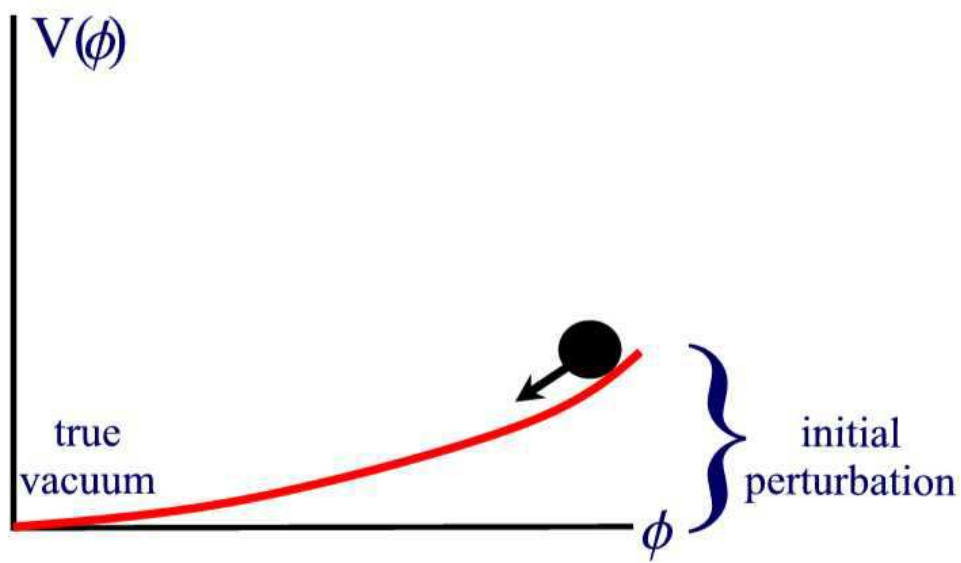}
\caption{Toy Model For Chaotic Inflation -- The inflaton finds
itself displaced from the true vacuum and proceeds to `roll' back.
Inflation takes place while the inflaton is displaced. }
\label{chaosinfl}
\end{figure}

\begin{figure}[]
\includegraphics[totalheight=5.0 in,keepaspectratio]{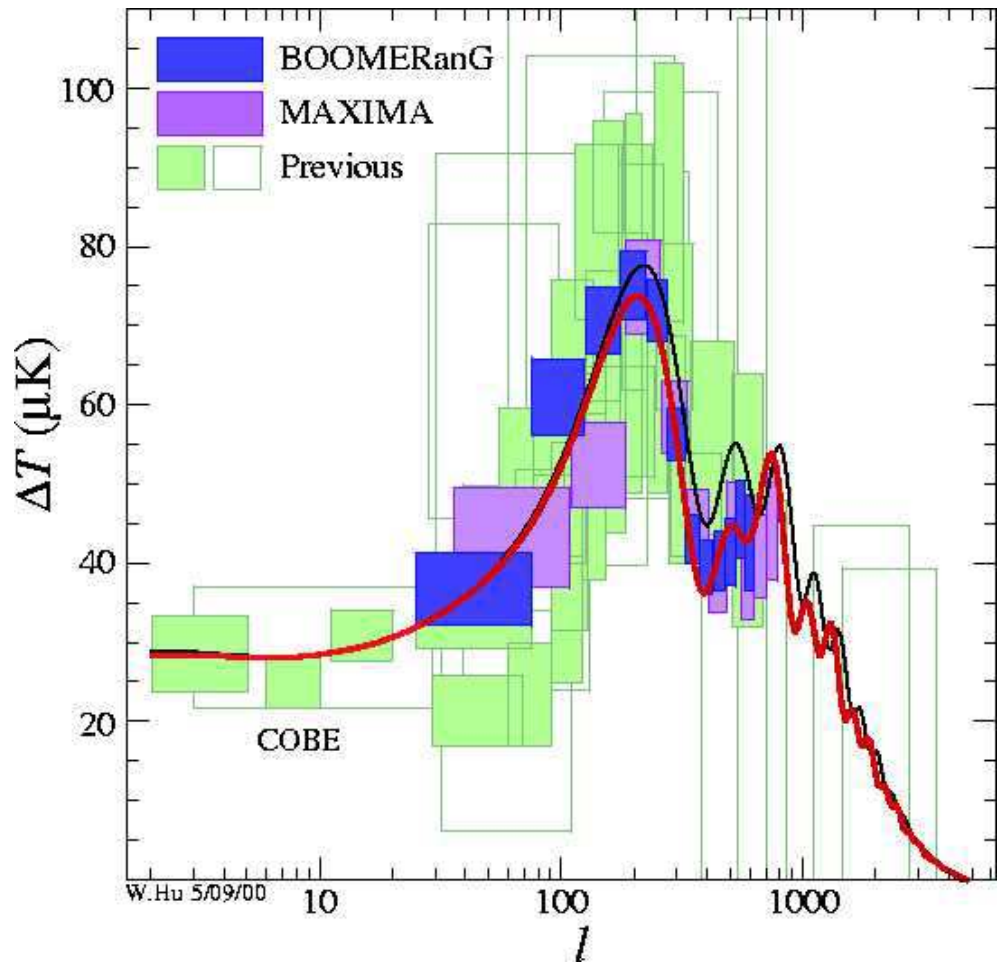}
\caption{Power Spectrum Plot of the Cosmic Background -- The data
is normalized to the quadrupole ($l=2$) anisotropy detected by
COBE.  The Doppler peak corresponds to a maximum power fluctuation
at a multipole of $l=220$, which is about an angle of one degree
on the sky.  This graph provided by Wayne Hu \citep{huhuhu}.}
\label{cmb}
\end{figure}
%------------------------------End Figures------------------------------------
\newpage
%-----------------------------Begin Tables------------------------------------
\begin{table}[h]
\caption{Summary of Friedman Models}
\begin{tabular}{cccl}
\hline

{\bf Density $\rho$}&\hspace{0.2in} {\bf Pressure $p$} &
\hspace{0.2in}{\bf $\alpha$} & \hspace{0.2in}{\bf Epoch} \\ \hline
\hline

$\rho_{R}$ &\hspace{0.2in} $\frac{\rho}{3}$ &
\hspace{0.2in}$\frac{1}{3}$ & \hspace{0.2in}Radiation Dominated \\

$\rho_{M}$ &\hspace{0.2in} 0 &\hspace{0.2in} 0 &
\hspace{0.2in}Matter Dominated (Non-relativistic Dust)  \\

$\rho_{\Lambda}$ &\hspace{0.2in} $-\rho$ &\hspace{0.2in} -1
&\hspace{0.2in} Vacuum Domination
\\

\hline
\end{tabular}
\label{table1}
\end{table}

\begin{table}[h]
\caption{Cosmological Models of a Matter Dominated Universe}
\begin{tabular}{lllll}

\hline

{\bf Geometry} \hspace{.4in} & {\bf $\Omega$} \hspace{.4in} & {\bf
$q_{0}$} &\hspace{0.2in} {\bf Fate of Universe} &
\hspace{0.2in}Name \\

\hline \hline

 Flat & $=1$ & $\frac{1}{2}$ &\hspace{0.2in} Open Universe &\hspace{0.2in}
Einstein-DeSitter Model
\\ Hyperbolic & $<1$ & $ < \frac{1}{2}$ &\hspace{0.2in} Open Universe&\hspace{0.2in}Open Model
\\ Spherical& $>1$ & $> \frac{1}{2}$ & \hspace{0.2in}Closed Universe&\hspace{0.2in}Closed
Model\\ \hline
\end{tabular}
\label{table2}
\end{table}
%------------------------------End Tables-------------------------------------
\end{document}